\documentclass{article}
\usepackage[hidelinks]{hyperref}
\usepackage[a4paper]{geometry}
\usepackage{xcolor}
\usepackage{verbatim}
\usepackage{graphicx}
\usepackage{stix2}
\usepackage{amsmath}
\usepackage{bm}
\usepackage{mathtools}
\usepackage[normalem]{ulem}
\usepackage{chemfig}
\usepackage[backend=biber,style=numeric-comp,giveninits=true,sorting=none,maxbibnames=30,doi=true,isbn=false,url=false]{biblatex}
\AtEveryBibitem{
    \clearfield{month}
    \clearfield{day}
    \clearfield{issn}
    \clearfield{isbn}
    \clearfield{endday}
    \clearfield{endmonth}
    \clearfield{issue}
    \clearfield{number}
    \clearfield{date}
    }
\renewbibmacro*{date}{
  \printfield{year}%
}

\renewbibmacro{in:}{}
\AtEveryBibitem{%
  \iffieldundef{eprint}
    {}
    {%
      \ifboolexpr{
        test {\iffieldequalstr{eprinttype}{arxiv}} or
        test {\iffieldequalstr{eprinttype}{arXiv}} or
        test {\iffieldequalstr{archiveprefix}{arxiv}} or
        test {\iffieldequalstr{archiveprefix}{arXiv}}
      }{}{%
        \clearfield{eprint}%
      }%
    }%
}
\DeclareFieldFormat[article]{pages}{#1} 
\usepackage{braket}
\renewcommand{\Re}{\operatorname{Re}}
\renewcommand{\Im}{\operatorname{Im}}

\newcommand{\ii}{\mathrm{i}}
\newcommand{\erf}{\mathrm{erf}}
\newcommand{\keywords}[1]{\noindent \textbf{Keywords:} #1}

\DeclareMathOperator*{\argmin}{arg\,min}
\DeclarePairedDelimiter\norm{\lVert}{\rVert}
\DeclarePairedDelimiter\abs{\lvert}{\rvert}
\usepackage{authblk} 
\author[1]{Simon Elias Schrader}
\author[1]{Håkon Emil Kristiansen}
\author[3]{Aleksander P. Woźniak}
\author[2]{Ludwik Adamowicz}
\author[1]{Simen Kvaal}
\author[1]{Thomas Bondo Pedersen \thanks{\texttt{t.b.pedersen@kjemi.uio.no}}}

\affil[1]{Hylleraas Centre for Quantum Molecular Sciences, Department of Chemistry, University of Oslo, P.O. Box 1033 Blindern, N-0315 Oslo, Norway}

\affil[2]{Department of Chemistry and Biochemistry, University of Arizona, 1306 E University Blvd, Tucson, Arizona 85721-0041, United States}

\affil[3]{Faculty of Chemistry, University of Warsaw, Pasteura 1, 02-093 Warsaw, Poland}

\title{Rothe's Method for Quantum Dynamics in Atoms and Molecules with Gaussian Wavepackets}

\date{\today}

\begin{document}

\maketitle

\begin{abstract}
    Capable of capturing both bound and continuum quantum dynamics, Gaussian wavepackets are highly attractive basis functions for simulating laser-driven processes in atoms and molecules. Unfortunately, fully flexible Gaussian wavepackets are exceedingly challenging to propagate in a numerically stable manner within the framework of conventional time-dependent variational principles. In this chapter, we discuss the sources of the numerical issues and review an alternative approach, Rothe's method, that offers a route to improved numerical stability. Recent proof-of-concept simulations based on Rothe's method indicate that Gaussian wavepackets provide results on par with highly accurate grid-based methods for both electronic and rovibrational quantum dynamics, including ultrafast nonlinear processes that involve the continuum such as high-harmonic generation. Remarkably few Gaussian wavepackets are needed to achieve the high accuracy of grid-based approaches, indicating that further algorithmic developments and efficient implementations may enable efficient simulations of not only electronic and rovibrational phenomena but also fully coupled electronic-nuclear quantum dynamics with significantly reduced memory demands. We also point out remaining practical challenges, including matrix elements of the squared Hamiltonian and the treatment of Coulomb cusps.
\end{abstract}


\section{Introduction}
\label{section: introduction}

The time evolution of the state of a nonrelativistic molecular quantum system is determined by the Schrödinger equation: Given $\Psi_0$, find $\Psi(t)$ for $0 \leq t \leq t_\text{final}$ such that
\begin{equation}
    \label{eq:time-dependent Schrödinger equation, general form}
    \ii\partial_t\Psi(t) = \hat{H}(t) \Psi(t), \qquad \Psi(0) = \Psi_0.
\end{equation}
Here, $\Psi(t)$ is a wavefunction on a $D$-dimensional spatial domain, possibly multicomponent in order to account for, e.g., spin. The corresponding Hilbert space is denoted $\mathcal{H}$.
While we will consider a variety of effective models in this article, the case of primary interest is a molecular system of $N$ particles with positions $r_i$ and charges $q_i$, governed by the Coulomb Hamiltonian (using atomic units throughout),
\begin{equation}
    \label{eq:molecular_hamiltonian}
    \hat{H}(t) = \sum_{i=1}^N \frac{\hat{p}_i^2}{2m_i} + \sum_{i<j}^N \frac{q_iq_j}{\vert {r}_i - {r}_j \vert} + \hat{U}(t),
\end{equation}
where $\hat{{p}}_i$ and $m_i$ are the momentum and mass of particle $i$, respectively,
and $\hat{U}$ represents interactions with external forces such as time-dependent electromagnetic fields. 
In lieu of general analytical solutions, different approaches have been developed for computing approximate wavefunctions.

The most straightforward approach is to discretize the full molecular wavefunction directly, after separation of the center-of-mass motion.
This is referred to as non-Born-Oppenheimer or pre-Born-Oppenheimer theory and has only been applied to few-body systems
due to the inherent exponential scaling with the number of particles $N$. Most non-Born-Oppenheimer methods have been
applied to the calculation of stationary states \cite{Bubin_Adamowicz_ECG2013,Mitroy2013,Matyus2019},
with fewer applications to full-dimensional molecular quantum dynamics induced by electromagnetic fields.
Recent examples include the simulation of laser-induced alignment of the HD molecule in Ref. \cite{Adamowicz2022} and 
the investigation of strong vibrational effects on two-photon double ionization of H$_2$ in Ref. \cite{arteaga_strong_2024}
(although the nuclear rotational degrees of freedom were frozen in the latter, with the molecular axis
fixed relative to the laser polarization direction).

The most widely used approaches are based on either the adiabatic or non-adiabatic Born-Op\-pen\-heimer approximation \cite{Born1927,Born1954}
where the concept of potential-energy surfaces replaces the bare Coulomb interaction in the Hamiltonian, reducing computational costs
significantly and thus allowing for studies of quantum dynamics in larger molecular systems.
While non-Born-Oppenheimer methods necessarily treat electrons and nuclei
on the same quantum-mechanical footing, the Born-Oppenheimer-based approaches are designed to describe particular types
of quantum-dynamical phenomena. For example, the multiconfigurational time-dependent Har\-tree (MCTDH) \cite{Meyer2009}
and ab initio multiple spawning (AIMS) \cite{Ben-Nun2000} methods can efficiently simulate non-adiabatic quantum (or semiclassical)
dynamics on a (limited) number of potential-energy surfaces, including ultrafast pump-probe spectroscopic experiments and
photochemical processes. Potential-energy surfaces of ionized species can also be taken into account
within the Feshbach close-coupling scheme but, like full non-Born-Oppenheimer methods, is only applicable to few-body systems
due to high computational costs \cite{palacios_theoretical_2015}.

Ultrafast electronic phenomena that occur on time scales short enough that nuclear motion can be neglected are often modelled
by time-dependent electronic-structure theory with clamped nuclei \cite{li_real-time_2020}. Examples include charge migration \cite{cederbaum_ultrafast_1999} and interactions with attosecond and few-femto\-sec\-ond laser pulses \cite{nisoli_attosecond_2017}. In some cases,
such as high-harmonic generation (HHG) \cite{lewenstein_theory_1994}, it is even possible to replace the correlated many-electron dynamics
with a much simpler single-active-electron model \cite{kulander_time-dependent_1988,kulander_dynamics_1993}.

Clamping only some of the nuclei---typically all nuclei but (selected) protons---and describing collectively the electrons and
non-clamped nuclei on the same quantum-mechanical footing is an intermediate approach that aims to capture the
most important quantum effects of coupled electronic-nuclear motion \cite{thomas_protonic_1969,thomas_protonic_1970}.
The most widely used approach belonging to this class is the nuclear-electronic orbital (NEO) method \cite{pavosevic_multicomponent_2020},
which has recently been extended to the simulation of quantum dynamics \cite{zhao_real-time_2020}.

All of these models, from non-Born-Oppenheimer theory over adiabatic and non-adiabatic approximations to purely electronic methods,
inherit an important feature from the molecular Coulomb Hamiltonian in eq.~\eqref{eq:molecular_hamiltonian}:
The Coulomb potential admits both discrete bound states and a continuous spectrum. The latter is crucial for
processes such as ionization and chemical bond dissociation and must be supported by the chosen discretization of the
wavefunction in a computational model. At the same time, the discretization must allow for accurate descriptions of
bound dynamics. In short, the discretization must be adaptive such that the approximate wavefunction can move freely
in Hilbert space as dictated by the Hamiltonian, preferably with controlled error.

These requirements are met by spatial grid-based approaches such as finite-difference \cite{smith_finite_1985,natan_real-space_2008},
pseudospectral (also known as the discrete variable representation) \cite{Meyer2009,hochstuhl_time-dependent_2014},
B-splines \cite{bachau_applications_2001}, or multiwavelet \cite{dinvay_multiresolution_2025,dinvay_multiresolution_2025-1} methods.
While grid-based methods can provide near-exact simulations, they often scale exponentially with the number of degrees of freedom, both
in computer time and in memory demands. As an example, the non-Born-Oppenheimer simulations of the hydrogen molecule by
\citeauthor{arteaga_strong_2024} required more than $400$ compute nodes (more than $21\,000$ CPU cores) to complete
due to the huge memory requirements, despite neglecting the nuclear rotational degrees of freedom \cite{arteaga_strong_2024}.

\begin{figure}[h!]
    \centering
    \includegraphics[width=\linewidth]{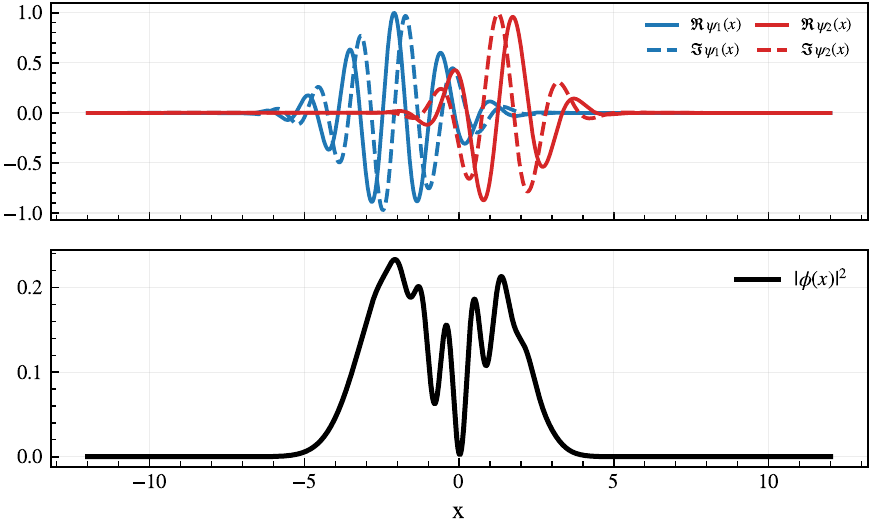}
    \caption{Superposition of two complex Gaussian wave packets: The component wave packets are $\psi_j(x)=\exp\!\left[-a_j(x-\mu_j)^2+\ii\,p_j(x-\mu_j)+\ii\gamma_j\right]$, for $j=1,2$, with parameters $a_1=0.22+0.08\ii$, $\mu_1=-2.1$, $p_1=4.2$, $\gamma_1=0.0$, and $a_2=0.34-0.05\ii$, $\mu_2=1.4$, $p_2=-3.1$, $\gamma_2=1.1$. The superposition is $\phi(x)\propto c_1\psi_1(x)+c_2\psi_2(x)$ with $c_1=1.0$ and $c_2=0.9$, and is normalized such that $\int |\phi(x)|^2\,dx=1$. Top panel: real and imaginary parts of $\psi_1$ and $\psi_2$. Bottom panel: normalized probability density $|\phi(x)|^2$.}
    \label{fig:superpos_two_gaussians}
\end{figure}

In this work, as a less memory-demanding and potentially more efficient alternative to grid-based approaches, 
we consider expansion of the wavefunction in a basis (Ga\-ler\-kin discretization) of Gaussian wavepackets (GWPs),
\begin{equation}
    \label{eq:GWP_def}
    \psi(x,t) = \exp\left[-(x-\mu)^TC(x-\mu) + \ii p^T (x-\mu) + \ii\gamma\right].
\end{equation}
For $D$ degrees of freedom, $x \in \mathbb{R}^D$.
Each GWP depends on time $t$ through the symmetric covariance matrix $C \in \mathbb{C}^{D \times D}$, the center
$\mu \in \mathbb{R}^D$, the momentum $p \in \mathbb{R}^D$, and the phase $\gamma \in \mathbb{C}$.
The condition $\Re(C) > 0$ guarantees $\psi \in L^2(\mathbb{R}^D)$, while
$\Im(\gamma)$ can be chosen such that the GWP is normalized.
When a set of normalized GWPs is used to expand a wavefunction (as in eq. \eqref{eq:gwp_expansion} below), $\Re(\gamma)$ can either be 
absorbed into the linear expansion coefficients ($c_i$ in eq. \eqref{eq:gwp_expansion} below) or, despite the obvious redundance, 
it can be kept in each GWP.
By expanding the wavefunction---either the many-body wavefunction directly or the single-particle functions used to
construct the many-body wavefunction---in GWPs with variable expansion length $M(t)$,
\begin{equation}
    \label{eq:gwp_expansion}
    \Psi(x,t) = \sum_{i=1}^{M(t)} \psi_i(x,t) c_i(t),
\end{equation}
it is formally possible to approximate the wavefunction with arbitrary accuracy at all times $t$ by adjusting the time-dependent
nonlinear Gaussian parameters, the linear expansion coefficients $c_i(t)$, and the expansion length $M(t)$ at each time step.
Figure \ref{fig:superpos_two_gaussians} illustrates the superposition of two complex Gaussians.

Importantly, the Hamiltonian matrix in GWP basis can be computed analytically, both for single-particle \cite{boys_electronic_1950}
and many-body Gaussians \cite{Boys1960,Singer1960}.

The GWPs form an overcomplete (\textit{vide infra}) basis of $L^2(\mathbb{R}^D)$ and
the extreme adaptivity of GWP expansions has recently been verified by fitting linear combinations of
GWPs to virtually exact
grid-based wavefunctions for two-dimensional models of electronic and of rovibrational quantum dynamics induced by rather extreme
time-dependent electric fields \cite{wozniak_gaussians_2024}. This study confirmed that both bound and unbound states with high angular momentum
can be accurately represented by relatively few Gaussians, provided that both linear and nonlinear parameters are allowed
to vary freely.

Furthermore, these observations indicate that not only purely electronic dynamics within the clamped-nuclei Born-Oppenheimer approximation \cite{Born1927,Born1954}
but also full-dimensional molecular quantum dynamics without
the Born-Oppenheimer (or Born-Huang \cite{Born1954}) approximation may be within reach by generalizing
explicitly correlated Gaussians (ECGs) \cite{Boys1960,Singer1960} to many-body Gaussian wave\-packets.
The ECGs have been used to calculate non-Born-Oppenheimer
stationary states of atoms and small mol\-e\-cu\-les,
yielding extremely accurate predictions of rovibrational spectroscopic observables, especially
when the leading-order relativistic and quantum electrodynamical corrections are included \cite{Bubin_Adamowicz_ECG2013,Mitroy2013,Matyus2019}.
Thus, it is not unreasonable to expect that many-body
GWPs may yield similar accuracy for ultrafast spectroscopic observables, provided that the full
non-Born-Oppenheimer molecular wavefunction can be propagated in a numerically stable manner.

Using GWPs as basis functions for quantum dynamics is not a new idea. Noting that GWPs are exact solutions for quadratic (harmonic)
potentials \cite{tannor2008introduction,heller2018semiclassical},
\citeauthor{heller1975time} developed a semiclassical theory of dynamics
in nonseparable (anharmonic) potentials \cite{heller1975time}, where a single GWP (its center and momentum) moves along the classical
trajectory but is allowed to stretch and rotate in phase space, thus capturing some quantum effects. In this
\textit{thawed Gaussian approximation}, the nonlinear parameters are determined in the local harmonic approximation,
which is obtained by approximating the potential by its second-order Taylor expansion about the current center of the
Gaussian. The thawed Gaussian approximation can be generalized to linear combinations of GWPs,
with linear and nonlinear parameters determined by the variational principle. The resulting equations of motion, even within
the local harmonic approximation, are more easily integrated if the widths of the Gaussians are kept fixed (the \textit{frozen
Gaussian approximation} \cite{Heller1981}). These and closely related ideas have been very successful in predicting, e.g.,
the vibrational structure of molecular electronic spectra---see \citeauthor{Vanicek2020} for a review \cite{Vanicek2020}.
Also in the AIMS method mentioned above, frozen GWPs moving along classical trajectories are used as basis functions, with
an algorithm deciding when and how to expand the basis with new GWPs. A recurring issue in these GWP-based approaches
is the numerical stability of the propagation of the nonlinear Gaussian parameters \cite{sawada1985strategy,kay1989matrix,wu_matching-pursuit_2003,habershon_linear_2012,koch_basis_2013,vMCG,lee_solving_2018,rowanSimulationHydrogenAtom2020,garashchuk2025variational,feischlRegularizedDynamicalParametric2026}.
The variational multiconfigurational Gaussian (vMCG) \cite{vMCG} method is inspired by the MCTDH approach and aims to
provide reliable non-adiabatic quantum dynamics for larger systems. Instead of expansions in single-particle functions,
the most general formulation of the vMCG approach uses linear combinations of multidimensional GWPs, with linear and nonlinear parameters determined variationally.
In the vMCG method, numerical stability issues are handled by various forms of regularization combined with
frozen and separable GWPs (in which case the vMCG ansatz reduces to an expansion in single-particle functions,
as in the MCTDH meth\-od) \cite{vMCG}.

In molecular electronic-structure theory \cite{Szabo1989,helgaker2013molecular},
specializations of single-particle GWPs are routinely used to expand molecular orbitals: Gaussian-type orbitals
(GTOs) \cite{boys_electronic_1950,shavitt_history_1993,Jensen2013,Nagy2017}.
The GTOs are time-independent functions with scalar real widths, fixed centers located at the clamped atomic nuclei, and zero momenta.
London orbitals \cite{London1937} (also known as gauge-including atomic orbitals)
have nonzero momenta related to an external magnetic vector potential and are used mainly to 
ensure gauge-origin invariance in calculations of perturbative and nonperturbative magnetic
properties \cite{tellgren_nonperturbative_2008,lange_paramagnetic_2012,helgaker_recent_2012}.
Basis sets of atom-centered GTOs are optimized to describe ground-state energies, with diffuse GTOs added to also
capture bound excited states. The continuum is not properly represented, though,
making conventional GTO basis sets unsuitable for simulating the quantum dynamics induced
by ultrashort and intense laser pulses where ionization processes are virtually unavoidable \cite{nisoli_attosecond_2017,palacios_quantum_2020,li_real-time_2020,ofstad_time-dependent_2023}.
Adding GTOs that mimic continuum functions \cite{kaufmannUNIVERSALGAUSSIANBASISSETS1989} have been used in combination with complex absorbing potentials
(CAPs) or the closely related heuristic lifetime model to compute time-dependent ionization probabilities, above-threshold ionization, and HHG
spectra \cite{Klinkusch2009,Sonk2011,Sonk2011a,luppi2012computation,luppi2013role,Krause2014,white2016computation,coccia2016gaussian,labeyeOptimalBasisSet2018,pauletti2021role,wozniakSystematicConstructionGaussian2021,Wozniak2022,coccia2022time,durden_evaluation_2025}.
While reasonable accuracy can be achieved with such extended GTO basis sets for specific observables such as HHG (see, e.g., \cite{coccia2016gaussian,labeyeOptimalBasisSet2018,wozniakSystematicConstructionGaussian2021}), it would clearly be preferable to use adaptive basis functions to capture the dynamics induced by one or more laser fields regardless of the chosen laser parameters (peak intensity, polarization, duration, envelope, carrier frequency, carrier-envelope phase, chirp rate, electric-dipole approximation or beyond).
To this end, GWPs may provide an attractive alternative.

The purpose of the present work is to review recent efforts aimed at numerically stable and accurate propagation of quantum-mechanical wavefunctions expressed
as linear combinations of GWPs. Section \ref{section: theory} provides the theoretical framework, including a discussion of conventional time-dependent variational
principles and their shortcomings, which serve as the motivation for introducing Rothe's method for solving the time-dependent Schrödinger equation using GWPs.
Our implementation of the aproach is described in Sec. \ref{sectionn: implementation} and proof-of-concept simulations from recent publications are summarized
in Sec. \ref{sec:applications}. Concluding remarks are found in Sec. \ref{section: conclusion}.

\section{Theory}
\label{section: theory}

\subsection{The time-dependent Schrödinger equation}

Our goal is to solve the time-dependent Schrödinger equation~\eqref{eq:time-dependent Schrödinger equation, general form} with a time-dependent Hamiltonian on the form
$$
    \hat{H}(t) = \hat{T} + \hat{V} + \hat{U}(t),
$$
where $\hat{T}$ is a kinetic energy operator, $\hat{V}$ is a static potential, and where $\hat{U}(t)$ is a time-dependent perturbation that typically models interactions with external electromagnetic fields. In Sec.~\ref{sec:applications} we consider a variety of effective models of the full Coulomb problem~\eqref{eq:molecular_hamiltonian}, from a simple 1D soft-Coulomb potential, via the 2D and 3D hydrogen atoms and the Henon--Heiles model, to nuclear dynamics in a Morse potential. The aim is a high-accuracy solution on a sufficiently dense temporal grid, capturing dynamics that are not merely small oscillations close to a stationary state, but complicated dynamics that involve buildup of highly detailed wavefunction components. In particular, the dynamics may have unbound parts, e.g., ionization and dissociation. Dynamics may happen at several distinct timescales due to, e.g., varying particle masses, a broad range of bound-state energy spacings, or resonances.

We are thus faced with an extremely challenging general computational problem. In order to retain as much accuracy as possible, one must carefully consider both which discrete representation of the wavefunction should be used, and also how it should be propagated in time. The particular choices must be tuned to the specific needs on a case-by-case basis, likely requiring compromises.

\subsection{Nonlinear ansatz manifolds}

In general there is a need for a \emph{nonlinear} approximation of $\Psi(t)$. Such are common in quantum chemistry: from the Hartree–Fock method via multiconfigurational self-consistent field approaches to coupled-cluster theory \cite{helgaker2013molecular}, the wavefunction is parameterized in a nonlinear fashion with an ansatz space $\mathcal{M}$, allowing the relevant section of Hilbert space $\mathcal{H}$ to be approximated.

The prevailing methodology for time-evolution of nonlinear (and linear) approximations views the local parts of ansatz space as \emph{smooth submanifolds $\mathcal{M}$ embedded in $\mathcal{H}$}. A smooth real (complex) embedded submanifold of dimension $n$ of a Hilbert space is a subset $\mathcal{M}\subset \mathcal{H}$ that locally ``looks like'' $\mathbb{R}^n$ ($\mathbb{C}^n$). For a general introduction to differentiable manifolds in physics, see, e.g., Refs.~\cite{carmoRiemannianGeometry1993,marsdenManifoldsTensorAnalysis2001}. 

We will here assume that the ansatz space is defined in terms of a set of \emph{coordinates} $y = (y^1,\cdots,y^n) \in Y \subset \mathbb{R}^n$  giving the wavefunction a \emph{parametric} dependence, i.e.,
$$
    \Psi = \Psi(x;y),
$$
where $x$ generically denotes the wavefunction's degrees of freedom, and is often omitted for ease of reading and writing (as was done in Sec.~\ref{section: introduction}). We focus for simplicity on real manifold coordinates $y$, since a complex coordinate can be viewed as two real coordinates.  At least locally near a point $\Psi \in \mathcal{M}$, the corresponding coordinates are unique, $y = y(\Psi)$, and they vary smoothly with $\Psi$. This is the essence of $\mathcal{M}$ (locally) being a smooth manifold.

An important mathematical construction is the \emph{tangent space} $T_\Psi \mathcal{M}$ of the manifold at the point $\Psi \in \mathcal{M}$. The tangent space consists of all allowed infinitesimal variations in $\Psi$, and is an $n$-dimensional real-linear space with basis $t_\mu(y) = (\partial/\partial y^\mu) \Psi(y)$. (This is what it means for $\mathcal{M}$ to ``locally look like $\mathbb{R}^n$''.) Thus, the tangent space consists of elements of Hilbert space on the form
\begin{equation}
    \delta \Psi = \sum_{\mu=1}^n \delta y^\mu t_\mu(y), \quad \delta y = (\delta y^\mu) \in \mathbb{R}^n.
\end{equation}
The tangent space is a real inner product space with inner product $\Re\braket{\delta\Psi_1|\delta \Psi_2}$, inherited from the ambient space $\mathcal{H}$.

\subsection{Method of vertical lines: The classical time-dependent variational principles}

Given  $\mathcal{M}\subset\mathcal{H}$ and an initial condition $\Psi(t=0) \in \mathcal{M}$, we seek an approximate time-evolution $\Psi(t) \in \mathcal{M}$ which in some way is optimal. To every $\Psi(t)\in\mathcal{M}$ we have coordinates $y(t) \in \mathbb{R}^n$ which form the unknowns of the approximation. For simplicity in this section, we will assume that $\mathcal{M}$ \emph{contains rays}: $\Psi\in\mathcal{M}$ if and only if for every nonzero $\alpha \in \mathbb{C}$, we have $\alpha\Psi \in \mathcal{M}$. This is the case for linear combinations of nonlinearly parameterized basis functions such as studied in this work.

Taking the time derivative and using the chain rule, we find
$$
    \dot{\Psi}(t) = \sum_\mu \dot{y}^\mu t_\mu(y(t)).
$$
If the TDSE is satisfied, $\dot{\Psi}(t) + \ii \hat{H}(t)\Psi(t)=0$. However, it is unlikely that $\ii\hat{H}(t)\Psi(t) \in T_\Psi\mathcal{M}$ in general, and it is necessary to introduce an approximation to force the derivative to lie in tangent space. There are two main approaches:
\begin{enumerate}
    \item The McLachlan variational principle (MVP) \cite{mclachlan_variational_1964} minimizes the norm of the residual error in the TDSE at any given time:
    \begin{equation}
        \dot{\Psi}(t) = \argmin_{\chi \in T_\Psi\mathcal{M}} \|\chi + \ii\hat{H}(t)\Psi(t)\|. \label{eq:MVP minimization}
    \end{equation}
    Equivalently, the residual is always orthogonal to tangent space.
        \item The Kramer--Saraceno variational principle (KSVP), often simply called the time-dependent variational principle, is a principle of least action~\cite{goldsteinClassicalMechanics2008} with action functional
\begin{equation}
    \mathcal{S}[\Psi(\cdot)] = \int_0^{t_\text{final}} \braket{\Psi(t)| \dot{\Psi}(t) + \ii \hat{H}(t) \Psi(t)} \, dt.
\end{equation}
 Informally, one requires $\delta \mathcal{S} = 0$ under infinitesimal variations of the history that preserve the endpoints $\psi(0)$ and $\psi(t_\text{final})$.
\end{enumerate}

The KSVP and the MVP are in general not equivalent, unless tangent space at $\Psi(y)$ is such that $\delta\Psi \in T_\Psi\mathcal{M}$ implies $\ii\delta\Psi \in T_\Psi\mathcal{M}$. Since tangent space is a \emph{real} vector space, this is not automatically true. It \emph{is} true if the variables $y^\mu$ are real and imaginary parts of complex variables, i.e., if the manifold $\mathcal{M}$ is actually parameterized with complex variables~\cite{lubich2008quantum}.

In the following, we focus on the MVP since it is most closely related to Rothe's method. Working out the minimization problem in Eq.~\eqref{eq:MVP minimization}, we obtain the following variational equation:
\begin{equation}
   \sum_\nu \Re \braket{t_\mu(y)|t_\nu(y)}\, \dot{y}^\nu = \Re \braket{t_\mu|-\ii\hat{H}(t)\Psi(t)}, \label{eq:MVP local coordinates}
\end{equation}
which is an implicit ODE system. The coefficient matrix $G(y)_{\mu\nu} = \Re \braket{t_\mu|t_\nu}$, being the matrix representation of the Riemannian metric on $\mathcal{M}$, must be inverted to evaluate the time derivative. 

The MVP is a quite natural approach to take when one seeks approximate dynamics, and has the interpretation that $-\ii\hat{H}(t)\Psi$ is projected orthogonally onto $T_\Psi\mathcal{M}$, providing local optimality of the approximation. However, it is quite typical that $G(y)$ beomes ill-conditioned, as in the example of linear combinations of explicitly correlated Gaussians, which means that even though the evolution is locally optimal, the ODE system can still be very stiff. This is associated with rapid change of the orthogonal projection operator onto $T_\Psi\mathcal{M}$ as $\Psi$ moves on the manifold, i.e., high or varying \emph{curvature} of the manifold.

\begin{figure}
    \centering
    \includegraphics[width=0.5\linewidth]{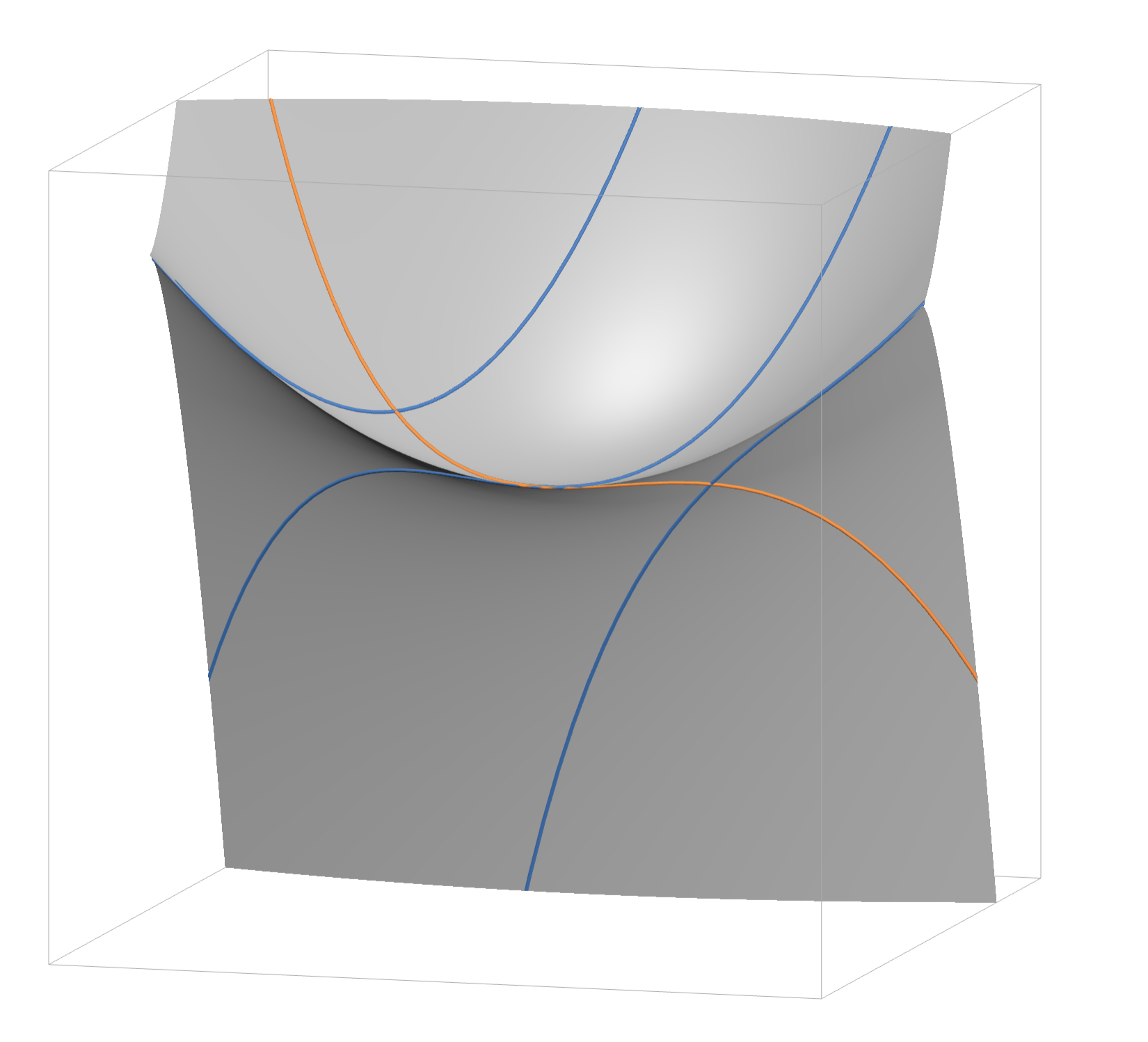}
    \caption{Illustration of a two-dimensional manifold with a singular line of dimensionality reduction. See text for details.}
    \label{fig:manifold example}
\end{figure}

In Figure~\ref{fig:manifold example}, a piece of a 2D parametric surface given by $\vec{r}(u,v) = (u, v^2, u^2 + 4v^3)$ is shown. The surface is an illustration of what can happen with a parameterized wavefunction $\Psi(x;y)$. The surface is not globally a smooth manifold. Instead, it is the union of 3 manifolds: two 2D pieces, and a 1D piece. The 1D piece is a singularity, in that the tangent basis vectors $\partial_u \vec{r}$ and $\partial_v \vec{r}$ become parallel there. Tangent space collapses to a 1D space, and  the $2\times 2$ matrix $G(u,v)$ is then singular. Also shown are curves that map orthogonal lines in coordinate $(u,v)$ space onto the surface. At the singular line, one can clearly see that the lines become parallel, illustrating the collapse of tangent space.

\subsection{Linear combinations of explicitly correlated Gaussians}

\begin{figure}
    \centering
    \includegraphics[width=\linewidth]{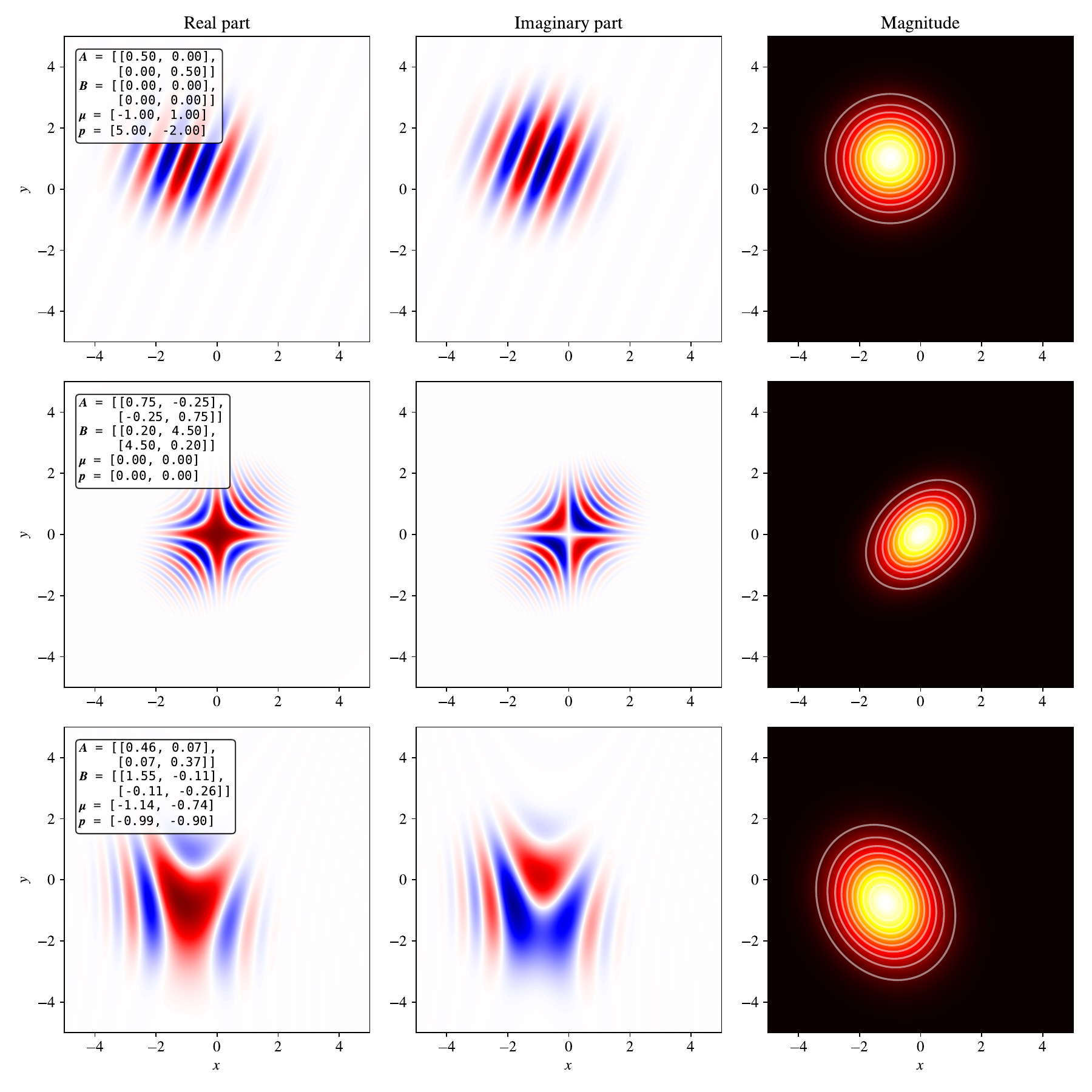}
    \caption{A few examples of Gaussians in $D=2$ dimensions, showcasing their qualitative behavior. For the real and imaginary parts, red colors correspond to large positive values, blue colors correspond to large negative values, while white is zero. All color ranges are normalized to the maximum absolute value of the Gaussian.}
    \label{fig:ECG examples}
\end{figure}

The primary ansatz space $\mathcal{M}$ in this work is a linear combination of complex-valued $D$-dimensional Gaussians. A complex Gaussian in $D$ dimensions is the exponential of a negative definite quadratic polynomial in $D$ variables. Such a polynomial may be parameterized in a number of ways. The perhaps most convenient form is by specifying the mean position $\mu \in \mathbb{R}^D$, the mean momentum $p\in\mathbb{R}^D$, and a complex symmetric covariance matrix $C = A + \ii B$ with $A > 0$, and define
\begin{equation}
    \gamma(x;\lambda) = \mathcal{N} \exp\left[ -(x - \mu)^\top (A + \ii B)(x-\mu) + \ii p^\top (x-\mu)\right], \label{eq:multidimensional gaussian}
\end{equation}
where $\mathcal{N}$ is a normalization constant that depends only on $A$,
$$
\mathcal{N} =
\left(\frac{2^D\det A}{\pi^D}\right)^{1/4}.
$$
Some Gaussians in $D=2$ dimensions are illustrated in Figure~\ref{fig:ECG examples}.

Now, $(A,B,\mu,p)$ constitute in total $D (D+3)$ free parameters $\lambda \in \mathbb{R}^{D(D+3)}$ in this Gaussian, and for a collection of $M$ Gaussians, we have in total $MD(D+3)$ independent parameters. However, it may be convenient to devise a possibly smaller number of independent parameters, say $\alpha \in \mathbb{R}^{N_\alpha}$, and let $\lambda_m=\Phi_m(\alpha)$ for some map $\Phi=(\Phi_1,\cdots,\Phi_M)$. We write
$$
    g_m(x) = \gamma(x;\Phi_m(\alpha)),
$$
for the $m$'th Gaussian thus parameterized. As an example of the use of $\Phi$, consider a case where one needs to force Gaussians to be real, central symmetric, and centered at the origin $x=0$. Then, $B_m=\mu_m=p_m=0$, and one may set $A_m = e^{\alpha_m} I$, with $\alpha = (\alpha_1,\cdots,\alpha_M)$ being the independent parameters; a single real-valued parameter per Gaussian.  Another example of the use of $\Phi$ is to enforce invariance under symmetry groups, by letting the parameters $\alpha$ parameterize multiple Gaussians in order to span an irreducible representation.

We refer to the Gaussian $g_m$ as an \emph{explicitly correlated Gaussian} (ECG) when $A+\ii B$ is not forced to be diagonal. When the $A + \ii B$ is \emph{a priori} known to be diagonal, we say that the Gaussian is \emph{uncorrelated}, which leads to some simplifications in analytical matrix elements derived in Section~\ref{sectionn: implementation}. (We remark that some authors reserve the term ``explicit correlation'' for the case when the $A + \ii B$ is block diagonal when the $D$ coordinates are associated with $N$ particles in $D/N$ spatial dimensions.)

The $M$ Gaussians form an $\alpha$-dependent linear subspace of Hilbert space, and we correspondingly consider arbitrary linear combinations:
\begin{equation}
    \Psi(x;c,\alpha) \equiv \sum_{m=1}^{M} c_m \, g_m(x;\alpha). \label{eq:linear combination of gaussians 1}
\end{equation}
Here, $c = (c_1,\dots,c_M)^\top$ are complex expansion coefficients. The ansatz~\eqref{eq:linear combination of gaussians 1} forms, at least locally, a nonlinear ansatz manifold $\mathcal{M}$, with local coordinates $y = (c,\alpha)$.

A word on indistinguishable particles is in order. In applications where multiple particles of the same or different species are treated, the wavefunction must be properly permutation (anti)symmetric. In the ECG community, such (anti)symmetry is usually handled using a spin-free method \cite{matsen_spin-free_1964}, premultiplying the wavefunction with a Young projection operator, which are particular linear combinations of particle index permutation operators. These could potentially be handled by the map $\Phi$ as well as constraints on $c$. However, this can quickly increase the number of Gaussians to an intractable number. It may be better to explicitly use the Young projection on an ansatz with fewer Gaussians in order to avoid computational overhead~\cite{Bubin_Adamowicz_ECG2013,hamermeshGroupTheoryIts1989}.

\subsection{Gaussians are extremely flexible}

We will now argue how linear combinations of Gaussians form an extraordinarily flexible ansatz, and are very well suited to describe a general time-dependent wavefunction. We argue in 1D for simplicity, but our conclusions hold for arbitrary dimensions. Consider the 1D harmonic oscillator with Hamiltonian $$\hat{H}_\text{HO} = -\frac{1}{2}\frac{d^2}{dx^2} + \frac{1}{2}\zeta^2 x^2.$$ The \emph{coherent states} are eigenfunctions of the ladder operator 
$$
\hat{a} = \frac{1}{\sqrt{2}}\left(\sqrt{\zeta} x + \frac{1}{\sqrt{\zeta}} \frac{d}{dx}\right),
$$
and are shifted Gaussians with a plane-wave factor,
\begin{equation}
    \varphi(x;p,q) = \braket{x|p,q} = \left(\frac{\zeta}{\pi}\right)^{1/4} e^{\ii pq/2} \exp\left[-\frac{1}{2}\zeta (x-q)^2+\ii p(x-q) \right]. \label{eq:coherent state}
\end{equation}
The eigenvalue of $\hat{a}$ is $z = \frac{1}{\sqrt{2}}\left(\sqrt{\zeta} q + \ii\frac{1}{\sqrt{\zeta}} p\right)$. We have chosen a parameterization of the coherent states that closely resembles the multidimensional Gaussian~\eqref{eq:multidimensional gaussian}.

The coherent states are complete in the sense that the identity operator over $L^2(\mathbb{R})$ can be written as an integral that, not coincidentally, resembles a phase-space integral:
\begin{equation}
    \hat{I} = \frac{1}{2\pi} \int_{\mathbb{R}^2} dp\,dq \, \ket{p,q}\bra{p,q}. \label{eq:completeness of coherent states}
\end{equation}
Here, $\ket{p,q}\bra{p,q}$ is the orthogonal projector onto $\varphi(x;p,q)$. Indeed, the integral indicates \emph{overcompleteness}: Applying the resolution to a coherent state $\varphi(x;q_0,p_0)$ shows that any coherent state is a linear superposition of coherent states with non-trivial expansion coefficients. Thus, one should be able to remove some coherent states and \emph{still} have a completeness relation. There are many such complete subsets of coherent states. For example, Perelomov has shown  \cite{perelomovCompletenessSystemCoherent1971} that if we select coherent states from a regular lattice $\{n_1 z_1 + n_2 z_2 \mid n_1,\, n_2 \in \mathbb{Z}\} \subset \mathbb{C}$ where the lattice vectors $z_1$ and $z_2$ are not parallel, and such that the area spanned by these vectors is $S \leq \pi$, then overcompleteness still holds. If $S=\pi$, one obtains a \emph{countable non-orthogonal basis} in the usual sense after removing a single arbitrary Gaussian from the lattice. The lattice expansions are quite intuitive when considered as trapezoidal quadrature approximations (which may be exact) of the integral \eqref{eq:completeness of coherent states}~\cite{bergoldGaussianWavePacket2024a}. 

Every exponent choice $\zeta$ is thus associated with an infinite set of lattices, each producing complete countable sets of Gaussians. It follows that any square integrable wavefunction $\psi(x)$ may be arbitrarily well approximated by \emph{finite} linear combinations of translated and scaled Gaussians in an \emph{infinite} number of possible ways, by selecting Gaussians with varying exponents and $(p,q)$-positions.

In quantum chemistry, \emph{even-tempered} sets of Gaussians are popular: Fix a \emph{single} arbitrary $(q,p)$-point, and instead select a sequence of \emph{exponents} according to the rule
\begin{equation}
    \zeta_n= u v^{n-1}, \quad n = 1,2,\ldots.
\end{equation}
\noindent Mathematically, one can show that under fairly mild conditions on the parameters $u$ and $v$ one obtains complete sets~\cite{klahnConvergenceRayleighRitzMethodb,shawCompletenessPropertiesGaussiantype2020,bachmayrErrorEstimatesHermite2014}. There is no reason to force the exponent to be real-valued: A complex-valued $\zeta$ adds complex quadratic oscillations to the Gaussian, which only adds to the flexibility for expanding wavefunctions.

To generalize these considerations to $D$ dimensions, consider the following: We may without loss of generality assume that $A$ in Eq.~\eqref{eq:multidimensional gaussian} is diagonal. We then obtain $\gamma(r,\lambda)$ as a tensor product of 1D Gaussians. We may thus generalize the resolution of identity~\eqref{eq:completeness of coherent states} to $D$ dimensions, and also the results that generate complete sets from discretizations of the integral. Moreover, the considerations on even-tempered sequences of exponents are not dimension-specific. Allowing $A$ to be non-diagonal just adds to the flexibility of the Gaussians.

We thus arrive at the conclusion that Eq.~\eqref{eq:linear combination of gaussians 1} is an extremely flexible nonlinear ansatz for the solution of the TDSE~\eqref{eq:time-dependent Schrödinger equation, general form}.

\subsection{Near-singularity of the MVP differential equations}

\begin{figure}
    \centering
    \includegraphics[width=.75\linewidth]{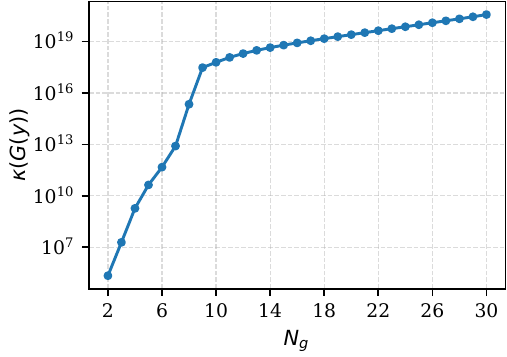}
    \caption{Condition numbers of the matrix $G(y)$ is shown for a 1D model of a single particle moving in a Gaussian potential, $V(x) = -\exp(-\mu x^2), \, \mu=0.1$, for the initial state $y=y(0)$. The initial basis is chosen as an even-tempered Gaussian basis set with exponents given by $\zeta_n = u v^{n-1}, \, n=1,\cdots,N_g$, with $u=2, v = \frac{10}{13}$ and $N_g$ being the number of Gaussians. The initial linear coefficients $\boldsymbol{c}$ are chosen as the lowest-energy eigenvector of the matrix of the Hamiltonian $H = -\tfrac{1}{2}\tfrac{\partial^2}{\partial x^2} + V(x)$.}
    \label{fig:condition numbers of gramian matrix}
\end{figure}

The variational MVP equations are an implicit ODE system on the form
$$
    G(y(t)) \dot{y}(t) = b(t,y(t)),
$$
where $G(y)_{\mu\nu} = \Re\braket{t_\mu|t_\nu}$ is the Gramian matrix of the tangent basis at $\Psi(x;y)$. Successful integration relies on the invertibility of $G(y)$ along the numerical trajectory. Unfortunately, $G(y)$
is often singular or ill-conditioned. Indeed, for linear combinations of ECGs, or any sufficiently flexibly nonlinearly parameterized basis, $G(y)$ will be severely ill-conditioned. For example, in Fig.~\ref{fig:condition numbers of gramian matrix}, the condition number of $G(y)$ for linear combinations of even-tempered Gaussians representing the
ground state in a 1D Gaussian potential is shown as a function of the number of Gaussians in the basis set.

We argue heuristically: Since $y = (c,\alpha)$, the tangent basis vectors come in two blocks:
$$
    t_m := \frac{\partial}{\partial c_m} \Psi(c,\alpha) = g_m, \quad t_{mj} := \frac{\partial}{\partial \alpha_{mj}} \Psi(c,\alpha) = c_m \frac{\partial}{\partial \alpha_{mj}} g_m.
$$
Note that the tangent space contains the basis vectors $g_m$, i.e., the Gaussians themselves. This is problematic: Suppose the parameterized basis $\{g_m(\alpha)\}_{m=1}^M$ is nearly complete. Each of the functions $t_{mj}$ is then very close to a linear combination of the $t_m = g_m$. Thus, the tangent basis is nearly linearly dependent. This implies that $G(y)$ will be nearly singular. Tangent space has a dimensional reduction, cf.~the discussion around Fig.~\ref{fig:manifold example}.

More generally, linear combinations of nonlinearly parameterized basis functions gives an ansatz space which is not a true manifold: Instead, the set $\mathcal{M}$ is the union of many pieces that are manifolds of varying dimension. At the intersection of these pieces, the tangent space does not have full dimension, and $G(y)$ becomes singular. This phenomenon is often referred to as \emph{collapsing tangent spaces}. For example, if only a single linear coefficient $c_m$ vanishes, $\Psi(x;y)$ becomes independent of $\alpha_m$, and in particular the dimension of tangent space is reduced, and $A(y)$ is singular. Similarly, if two basis functions become equal, the corresponding linear coefficients are non-unique, and the tangent space is again reduced in dimension.

This argument is also central to the discussion of instabilities and how they are handled numerically within the vMCG approach in Ref.~\cite{vMCG}. 
In a sense, the argument expresses the simple notion that if one expands the wavefunction in a complete basis, then there is no need to let the basis functions evolve in time.

This phenomenon also shows that, when using the MVP or the KSVP on an ansatz of the generic form ``linear combination of nonlinearly parameterized basis functions'', one can have a \emph{too flexible ansatz}. Counterintuitively, one will, in general, not get an improved time-dependent approximation by adding more parameterized basis functions.

It seems to be almost standard operating procedure to regularize the implicit system by selecting a small $\epsilon>0$ and make the replacement
\begin{equation}
    G(y) \leftarrow G(y) + \epsilon I \label{eq:regularization hack in MVP}
\end{equation}
in the MVP/KSVP. Although this regularization gives satisfactory results in many cases, e.g., for computing spectra via autocorrelation functions, it should be done with care, as it changes the equations being solved, and therefore in general breaks the accuracy of the time evolution~\cite{feischlRegularizedDynamicalParametric2026}.

\subsection{Method of horizontal lines: Rothe's Method}

\begin{figure}
    \centering
    \includegraphics[width=0.75\linewidth]{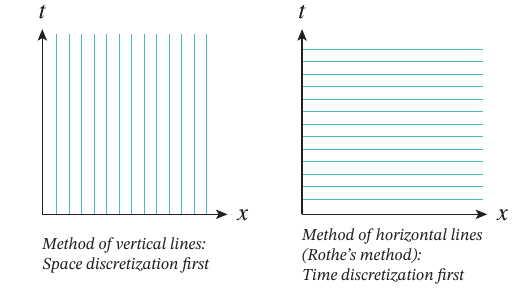}
    \caption{Illustration of the classical PDE discretization approach vs.~Rothe's method.}
    \label{fig:horizontal-vertical}
\end{figure}

\subsubsection{Rothe's Method for the Time-Dependent Schrödinger Equation}

Solving the MVP or KSVP equations of motion is probably near impossible for linear combinations of highly flexible basis functions. A method that alleviates this problem is Rothe's method \cite{Rothe1930_MA,deuflhard2012adaptive}, which is also referred to as the \emph{method of time layers}, \emph{method of horizontal lines}, \emph{transverse method of lines}, or \emph{method of lines transpose}.  Instead of discretizing the spatial dimensions first and then deriving ordinary differential equations for the equations of motion, Rothe's method starts by discretizing the time dimension first, thereby converting time evolution into a sequence of spatial problems that need to be solved, see Fig~\ref{fig:horizontal-vertical}. The method can be used both to derive analytical properties about PDEs, such as existence and uniqueness, as well as devise numerical schemes. The basic idea is that the spatial representation can be chosen independently for each time step.

In the context of quantum dynamics, Rothe's method has previously been used to solve the quantum-classical Liouville equation \cite{Horenko2004adaptive}, as well as for the time propagation of neural network quantum states \cite{Gutierrez2022}. In this work, we focus on the application of Rothe's method to propagate linear combinations of Gaussians, which has been demonstrated in Refs. \cite{kvaal2023need,schraderTimeEvolutionOptimization2024,wozniak2025rothetimepropagationcoupled,schraderMultidimensionalQuantumDynamics2025,schraderTimeDependentGaussianBasis2025}.

A semidiscrete version of Eq.~\eqref{eq:time-dependent Schrödinger equation, general form} can be devised using $A$-stable implicit Runge--Kutta methods~\cite{hairerSolvingDifferentialEquations} applied directly to the Hilbert space formulation of the TDSE~\eqref{eq:time-dependent Schrödinger equation, general form}. For simplicity, we consider the implicit midpoint method, or the single-stage Gauss--Legendre collocation method, which for the TDSE becomes the Crank--Nicolson method: Starting from the initial condition $\Psi(t_0)$, a sequence of Hilbert space vectors $\Psi(t_{i})$, with $t_i =  i \Delta t$, is defined by the relation
\begin{equation}
\label{eq:crank-nicolson-Schrodinger}
    \hat{A}_{i}\Psi(t_{i+1})- \hat{B}_{i}\Psi(t_i) = 0,
\end{equation}
where the operators are defined as
\begin{equation}
    \hat{A}_{i}=\hat I+\ii\frac{\Delta t}{2}\hat H\left(t_{i+1/2}\right), \ \hat{B}_i=\hat I-\ii\frac{\Delta t}{2}\hat H\left(t_{i+1/2}\right). \label{eq:definition of A and B}
\end{equation}
Here, $\hat I$ is the identity operator, and for simplicity we assume that the self-adjoint operator $\hat{H}(t)$ has a time-independent domain, i.e., the set of vectors $\Psi$ for which $\hat{H}(t)\Psi \in \mathcal{H}$. Since the spectrum of $\hat{H}(t)$ is real, $\hat{A}_i^{-1}$ exists and is continuous. Equation \eqref{eq:crank-nicolson-Schrodinger} thus gives an abstract time-stepping scheme preserving the domain of the Hamiltonian.

The idea behind Rothe's method is that each $\Psi(t_{i})$ may have an independent spatial approximation space $\mathcal{M}(t_i)$. We recast the abstract time stepping scheme as an optimization problem, seeking the optimal element in each ansatz space:
\begin{equation}\label{eq:Rothe}
    \Psi(t_{i+1})=\argmin_{\chi \in \mathcal{M}(t_{i+1})} \mathrm r^2_{i+1}(\chi),
\end{equation}
where the \emph{Rothe error} is defined as
\begin{equation}
    \mathrm r_{i+1}^2(\chi)=\norm{\hat{A}_i\chi-\hat{B}_i\Psi(t_i)}^2.
\end{equation}
Note that in contrast to the MVP/KSVP approach, there is no need to \emph{a priori} have a fixed ansatz space $\mathcal{M}$. Each time step may have an updated ansatz space. This notion of adaptivity is one of the main benefits of the Rothe method.

Equation~\ref{eq:Rothe} is closely related to the MVP principle defined in Eq.~\eqref{eq:MVP minimization}. Indeed, whereas the MVP attempts do resolve the optimal formulation of the TDSE in tangent space, Rothe's method attempts to find the manifold element that most closely approximates a \emph{finite time step} using the chosen time stepping scheme. The main implication is that Rothe's method is not \emph{a priori} hampered by singularities of the manifold $\mathcal{M}(t_i)$.

The Crank--Nicolson scheme is a useful method for the TDSE, as it is symplectic, energy conserving (for time-independent Hamiltonians), and norm conserving. Considering it is a second-order scheme and of relatively low complexity, it represents a reasonable compromise of accuracy and computational cost. We are not aware of any implementations of Rothe's method that goes beyond the Crank--Nicolson scheme in accuracy, although possible schemes could be the $s$-stage Gauss--Legendre method~\cite{hairerGEOMETRICINTEGRATIONORDINARY}.

\subsection{Rothe's method for explicitly correlated Gaussians}
\label{sec:rothe for gaussians}

As the next step, we apply the Gaussian ansatz~\eqref{eq:linear combination of gaussians 1} with $M(t)$ basis functions that may vary between time steps:
\begin{equation}\label{eq:WF_ansatz}
    \chi(t)=\sum_{m=1}^{M(t)} c_m(t)g_m(\alpha(t)) \in \mathcal{M}(t).
\end{equation}
The optimization in Eq.~\eqref{eq:Rothe} is now carried out with respect to the ansatz parameters $(\alpha,c)$, i.e.,
\begin{equation}\label{eq:Rothe_opt}
 \bigl(\alpha(t_{i+1}),c(t_{i+1})\bigr)=\argmin_{(\alpha,c)}\ \norm{\sum_{m=1}^{M(t_{i+1})}c_m \hat{A}_i g_m(\alpha)-\sum_{m=1}^{M(t_{i})}c_m(t_i) \hat{B}_i g_m(\alpha(t_i))}^2,
\end{equation}
where $c=(c_1,\dots,c_{M(t_{i+1})})^\top$, and where $\hat{A}_i$ and $\hat{B}_i$ are defined in Eq.~\eqref{eq:definition of A and B}.

\subsubsection{Variable Projection}
\label{sec:VarPro}

For a given set of nonlinear parameters $\alpha$, the optimal \emph{linear} coefficients $c$ can be found analytically using the \textit{Variable Projection} algorithm (VarPro) \cite{Golub_Pereyra}. That is, we can write $c$ as an explicit function of $\alpha$. 
Using VarPro, the optimal linear coefficients $c^{i+1}(\alpha)$ for a given $\alpha$ in eq.  \eqref{eq:Rothe_opt} are formally given by
\begin{equation}\label{eq:LinSolve}
    {c}^{i+1}({\alpha})= \left[\check{{S}}^{i+1}(\alpha)\right]^{-1}{\rho}^{i+1}( \alpha).
\end{equation}
Here, we introduced the weighted overlap matrix $\check{{S}}^{i+1}(\alpha)\in  \mathbb{C}^{M(t_{i+1})\times M(t_{i+1})}$ and the vector ${\rho}^{i+1}(\alpha)\in \mathbb{C}^{M(t_{i+1})}$ with elements
\begin{equation}
\begin{aligned}
        \check S^{i+1}_{mn}({\alpha})
    &= \braket{ g_m(\alpha) | \hat{A}^\dag_i \hat{A}_i| g_n(\alpha) } \\
    &= \braket{ g_m({\alpha}) | \hat I - \Im(\Delta t)\hat{H}\!\left(t_{i+1/2}\right) + \frac{\lvert{\Delta t}\rvert^2}{4} \hat H^2\!\left(t_{i+1/2}\right) | g_n({\alpha}) },
\end{aligned} \label{eq:weighted inner product gramian}
\end{equation} 
\begin{equation}
\begin{aligned}
        \rho^{i+1}_{m}({\alpha})
    &= \braket{ g_m(\alpha) | \hat{A}^\dag_i \hat{B}_i | \Psi(t_i) } \\
    &= \braket{ g_m({\alpha})| \hat I - \ii \Re(\Delta t )\hat{H}\!\left(t_{i+1/2}\right) - \frac{\lvert\Delta t\rvert^2}{4}\hat{H}^2\!\left(t_{i+1/2}\right) | \Psi(t_i) }. 
    \label{eq:weighted inner product rho}
\end{aligned}
\end{equation}
These expression show that one needs to compute matrix elements of $\hat{I}$, $\hat{H}(t_{i+1/2})$, and $\hat{H}^2(t_{i+1/2})$ between different Gaussian basis sets, the bra and ket side being the Gaussians $g_m(\alpha)=\gamma(x;\Phi_m(\alpha))$, and $g_n(\alpha') = \gamma(x;\Phi_n(\alpha'))$, with $\alpha' = \alpha(t_i)$ entering the ket side of the $\rho^{i+1}(\alpha)$ formula.

In these expressions, $\Delta t$ can be complex, allowing for both real and imaginary time propagation, if desired. The matrix $\check{{S}}^{i+1}(\alpha)$ is Hermitian. Equation~\eqref{eq:LinSolve} is the usual expression for finding the coefficients of a vector with respect to a non-orthogonal basis, here the $M(t_{i+1})$ weighed Gaussians $\hat{A}_i g_m$. We correspondingly define the orthogonal projector onto the weighted Gaussians,
\begin{equation}
\hat{P}_{i+1}(\alpha)
=
\sum_{mn}
\hat{A}_i |{g}_m(\alpha)\rangle
\left[\check  S^{i+1}( \alpha)\right]^{-1}_{mn}
\langle {g}_n( \alpha)| \hat{A}_i^\dag. \label{eq:weighted Projection operator}
\end{equation}
If we insert the analytical solution $c(\alpha)$ into  Eq.~\eqref{eq:Rothe_opt}, we obtain a Rothe error that depends only on ${\alpha}$:
\begin{equation} \label{eq:Rothe error abstract form}
     \mathrm{r}^2_{i+1}({\alpha})= \|(\hat{I}- \hat{P}_{i+1}(\alpha))\hat{B}_i\Psi(t_i)\|^2 .
\end{equation}
This expression has the interpretation that the Rothe error is the  error in the approximation of $\hat{B}_i\Psi(t_i)$ in the subspace spanned by the $M(t_{i+1})$ weighted Gaussians $\hat{A}_i g_m(\alpha)$.

The Rothe error can be expressed as
\begin{equation}\label{eq:RotheError}
\begin{aligned}
        \mathrm{r}^2_{i+1}({\alpha})
    &=\braket{\Psi(t_i)|\hat{B}_i^\dag \hat{B}_i|\Psi(t_i)}- \rho^{i+1}({\alpha})^\dagger \check{{S}}^{i+1}(\alpha)^{-1}\rho^{i+1}({\alpha})\\
    &=\braket{\Psi(t_i)|\hat{B}_i^\dag \hat{B}_i|\Psi(t_i)} - \rho^{i+1}({\alpha})^\dagger  c^{i+1}({\alpha}).
 \end{aligned}
 \end{equation}
The first term is independent of $\alpha$ and can be computed outside the optimization.

Optimizing the Rothe error gives the solution
\begin{equation}\label{eq:Rothe_alpha_only}
{\alpha}(t_{i+1})=\argmin_{{\alpha}}\left[\mathrm{r}^2_{i+1}({\alpha})\right],
\end{equation}
and
\begin{equation}
     c(t_{i+1})= c^{i+1}({\alpha}(t_{i+1})).
\end{equation}
Assuming that the final Rothe error is sufficiently small, this solution determines $\Psi(t_{i+1})$:
$$
    \Psi(t_{i+1}) = \sum_{m=1}^{M(t_{i+1})} c_m(t_{i+1}) g_m(\alpha(t_{i+1})).
$$
If the Rothe error is deemed unsatisfactory, the Gaussian basis can be augmented with a trial function and the Rothe error re-optimized to achieve a more accurate result. This is described further in Sec.~\ref{sec:BasisFlexibility}.

\subsubsection{Gradient of the Rothe error}
\label{sec:rothe gradient}

To minimize Eq. \eqref{eq:RotheError}, several optimization algorithms require its gradient with respect to $\alpha$. For each component $\alpha_k$, $k=1,\dots,N_{\alpha}(t)$, omitting the dependence of time step $i$ for notational convenience,
the partial derivative of the squared Rothe error reads
\begin{equation}
\begin{aligned}
        \frac{\partial \mathrm{r}^2(\alpha)}{\partial \alpha_{k}} &=-\frac{\partial}{\partial \alpha_{k}}\left(\rho(\alpha)^\dagger\check {S}^{-1}(\alpha) \rho( \alpha)\right)\\
        &=-\left(\frac{\partial  \rho(\alpha)}{\partial \alpha_{k}}^\dagger c( \alpha)
        - c( \alpha)^\dagger\frac{\partial \check{ S}( \alpha)}{\partial \alpha_{k}} c( \alpha)
        + c( \alpha)^\dagger\frac{\partial  \rho( \alpha)}{\partial \alpha_{k}}\right).
\end{aligned} \label{eq:gradient of squared Rothe error}
\end{equation}
Here, we used the identity
\begin{equation}
    \frac{\partial \check{ S}^{-1}( \alpha)}{\partial \alpha_{k}}=-\check{ S}^{-1}( \alpha)\frac{\partial \check{ S}( \alpha)}{\partial \alpha_{k}}\check{ S}^{-1}( \alpha).
\end{equation}
The partial derivatives in Eq.~\eqref{eq:gradient of squared Rothe error} are conveniently resolved by means of the chain rule. For the partial derivatives of $\rho(\alpha)$, we obtain
\begin{equation}
    \begin{split}
        &\frac{\partial}{\partial \alpha_k} \rho(\alpha)_m = \frac{\partial}{\partial \alpha_k}  \sum_{n=1}^{M(t_i)} \braket{g_m(\alpha)  | \hat{A}^\dag_i \hat{B}_i |  g_n(\alpha(t_i)) } c_n(t_i) \\
    &\,= \sum_{n=1}^{M(t_i)} \sum_{\ell=1}^{D(D+3)} \frac{\partial\Phi_{m\ell}(\alpha)}{\partial \alpha_k} \frac{\partial}{\partial \lambda_\ell} \braket{\gamma(\Phi_m(\alpha))  | \hat{A}^\dag_i \hat{B}_i |  \gamma(\Phi_n(\alpha(t_i))) } c_n(t_i).
    \end{split}
\end{equation}
The needed partial derivatives with respect to Gaussian parameters in the ``bra'' are conveniently obtained analytically or with automatic differentiation if the matrix element itself is known on closed form. The partial derivatives of the weighted overlap matrix $\check{S}(\alpha)$ are similarly obtained.

\subsection{Properties of Rothe's method for the TDSE}

\subsubsection{Relation of the Rothe error to the variance of the Hamiltonian}\label{sec:Rothe_variance}

The Rothe error is closely related to the component of $\hat{H}(t)\Psi(t)$ outside the variational space, and, for projected eigenstates, it reduces to the variance of the Hamiltonian. To see this, we consider the Rothe error for a time-independent Hamiltonian $\hat H$ and keep $\alpha$ fixed for simplicity. That is, we consider the Rothe error when optimizing only the linear coefficients. For notational convenience, we will omit $\alpha$.

Let $\mathcal V \subset \mathcal{H}$ be the finite-dimensional linear space spanned by the basis functions $\{g_m\}_{m=1}^{M(t)}$, and let $\hat P$ denote the orthogonal projector onto $\mathcal V$. Let $\Psi\in \mathcal V$ be the wave function at time $t$, which we assume to be normalized. The optimal squared Rothe error, going from time $t$ to $t+\Delta t$, is then
\begin{equation}
\mathrm r^2_{\mathrm{opt}}=\min_{\chi\in\mathcal V}\norm{\hat A\chi-\hat A^\dagger\Psi}^2,\qquad
\hat A=\hat I+\ii\frac{\Delta t}{2}\hat H.
\end{equation}
To first order in $\Delta t$, 
\begin{equation}
\chi=\Psi+\Delta t\,\eta + O(\Delta t^2),
\qquad \text{where }
\eta\in\mathcal V.
\end{equation}
This gives
\[
\hat A\chi-\hat A^\dagger\Psi
=
\Delta t(\eta+\mathrm i\hat H\Psi)
+
\mathrm i\frac{\Delta t^2}{2}\hat H\eta.
\]
To leading order in $\Delta t$, the minimization problem hence reduces to
\begin{equation}
\mathrm r_{\mathrm{opt}}^2=
\Delta t^2\min_{\eta\in\mathcal V}\norm{\eta+\ii\hat H\Psi}^2+O(\Delta t^3).
\end{equation}
The exact TDSE selects $\eta=-\ii\hat H\Psi$, corresponding to the first-order Euler approximation of the exact dynamics. Since $\eta$ is constrained to lie in $\mathcal V$, the optimal admissible choice is its orthogonal projection onto $\mathcal V$, which is identical to the solution obtained from the MVP~\eqref{eq:MVP minimization},
\begin{equation}
\eta_{\mathrm{opt}}=-\ii\hat P\hat H\Psi.
\end{equation}
Therefore,
\begin{equation}
\eta_{\mathrm{opt}}+\ii\hat H\Psi
=
\ii(\hat I-\hat P)\hat H\Psi,
\end{equation}
and the squared Rothe error becomes
\begin{equation}
\mathrm r_{\mathrm{opt}}^2=
\Delta t^2\norm{(\hat I-\hat P)\hat H\Psi}^2+O(\Delta t^3).
\end{equation}
Writing this as an inner product we get
\begin{align}
\mathrm r_{\mathrm{opt}}^2
&=
\Delta t^2\bra{\Psi}\hat H(\hat I-\hat P)\hat H\ket{\Psi}+O(\Delta t^3)\nonumber\\
&=
\Delta t^2\left(
\bra{\Psi}\hat H^2\ket{\Psi}
-
\bra{\Psi}\hat H\hat P\hat H\ket{\Psi}
\right)
+O(\Delta t^3).
\end{align}
If $\ket{\Psi}$ is an eigenvector of the projected Hamiltonian $\hat P \hat H \hat P$,
\begin{equation}
\mathrm r_{\mathrm{opt}}^2 =
\Delta t^2\left(
\bra{\Psi}\hat H^2\ket{\Psi}
-
\bra{\Psi}\hat H\ket{\Psi}^2
\right)
+O(\Delta t^3)
=
\Delta t^2\mathrm{Var}(\hat H;\Psi)+O(\Delta t^3).
\end{equation}

This dependence of the time evolution error on the variance of the Hamiltonian is also found in methods that use a time-dependent variational principle \cite{lubich2008quantum}.
The variance of the Hamiltonian (the energy variance, for time-independent cases) is a key quantity for highly accurate wavefunction optimization in electronic-structure theory 
because the squared Hamiltonian operator accentuates local errors in the wavefunction to a much greater extent than the Hamiltonian itself, especially at or near
the nuclei \cite{schwartz_lower_1967,keaveny_calculation_1969}.
The energy variance is crucial for establishing accurate \emph{lower bounds} of the ground-state energy \cite{ireland_lower_2022,ronto_lower_2023}
and it is frequenctly used in quantum Monte Carlo approaches \cite{foulkes_quantum_2001,toulouse_optimization_2007}.
Furthermore, it can be used to access excited states with the variational quantum eigensolver \cite{tilly_variational_2022,zhang_variational_2020}.

\subsubsection{The Rothe error as an upper bound for the time-evolution error}

For real-time propagation, the Rothe error is an upper bound for the time evolution error relative to exact Crank-Nicolson propagation \cite{schraderTimeEvolutionOptimization2024}.
Let $\tilde \Psi(t_{i+1})=\hat A_{i}^{-1}\hat A_{i}^\dagger{\Psi(t_i)}$ be the exact solution of the Crank-Nicolson propagation scheme and $\Psi(t_{i+1})$ an approximate solution. We have that
\begin{equation}
    \begin{aligned}
        \left\|\Psi(t_{i+1})-\tilde \Psi(t_{i+1})\right\|&= \left\|\Psi(t_{i+1})-\hat A_{i}^{-1}\hat A_{i}^\dagger{\Psi(t_i)}\right\|\\
        &\leq \left\|\hat A_{i}^{-1}\right\|\left\|\hat A_{i}\Psi(t_{i+1})-\hat A_{i}^\dagger{\Psi(t_i)}\right\|\\
        &\leq \mathrm r_{i+1}.
    \end{aligned}
\end{equation}
where we used that $\hat A_{i}^{-1}$ is normal and has spectrum inside the unit circle in the complex plane.
This result shows that the time evolution error going from $t_0$ to $t_n=t_0 + n\Delta t$ is at most 
\begin{equation}
    \norm{\Psi(t_n)-\tilde \Psi(t_n)}\leq \sum_{i=1}^n \mathrm{r}_i\equiv \mathrm{R}_n,
\end{equation}
where $\tilde\Psi(t_n)=\hat A_{n-1}^{-1}\hat A_{n-1}^\dagger\cdots \hat A_0^{-1}\hat A_0^\dagger \Psi(t_0)$ is the result of applying $n$ exact Crank-Nicolson steps to $\Psi(t_0)$, and $\mathrm{R}_n$ is the cumulative Rothe error at time $t_n$.

\subsection{Rothe's method for time-dependent mean-field methods}\label{sec:Rothe_MeanField}

Rothe's method is not restricted to the TDSE and can instead be applied to orbital time evolution equations. In time-dependent Hartree-Fock (TDHF) \cite{mclachlanTimeDependentHartreeFockTheory1964,goingsRealtimeTimedependentElectronic2017} theory and time-dependent Kohn-Sham density-functional theory (TDDFT) \cite{yabanaTimedependentLocaldensityApproximation1996,TDDFT,tancogne-dejeanOctopusComputationalFramework2020}, the $N$-electron wave function $\Psi({x}_1,\dots,{x}_N,t)$ is expressed as a Slater determinant consisting of $N$ time-dependent orbitals $\Phi(t)\equiv\{\varphi_j({x},t)\}_{j=1}^N$:
\begin{equation}\label{eq:SlaterDeterminant}
    \Psi({x}_1,\dots,{x}_N,t)={\sqrt{N!}}\mathcal {A} \left(\varphi_1( x_1,t)\dots\varphi_N( x_N,t)\right),
\end{equation}
where $\mathcal{A}$ is the antisymmetrization operator, and where $x_i = (r_i,s_i)$ are the space and spin degrees of freedom for electron $i$. The corresponding one-electron spatial density reads
\begin{equation}\label{eq:Density}
    \rho(r,t) = N \int \abs{\Psi(r, s_1,  x_2, \ldots,  x_N,t)}^2 \, \mathrm{d}s_1 \mathrm{d} x_2 \cdots \mathrm{d} x_N.
\end{equation}
In TDHF and TDDFT, time evolution of the wave function and the density, respectively, is expressed through orbital time evolution equations
\begin{equation}\label{eq:TDHFDFT}
    \frac{\partial}{\partial t} \varphi_j=-\ii \hat{h}(\Phi(t),t) \varphi_j, \quad j=1,\dots,N.
\end{equation}
In this expression, $\hat{h}(\Phi(t),t)$ represents the single-particle operator that governs the time evolution of each orbital. For TDHF, it is the Fock operator $\hat{h}(\Phi(t),t)=\hat{F}(\Phi(t),t)$, while for TDDFT, it is the kinetic energy operator plus the Kohn-Sham potential, $\hat{h}(\Phi(t),t)=-\dfrac{1}{2}\nabla^2+\hat{v}_s(\Phi(t),t)$. Equation \eqref{eq:TDHFDFT} can be discretized with time step $\Delta t$. Defining the operators
\begin{equation}
    \tilde A_{i}=\hat I+\ii \frac{\Delta t}{2}\hat h\left(\Phi(t_i),t_i+\frac{\Delta t}{2}\right), \qquad\tilde B_i=\hat I-\ii \frac{\Delta t}{2}\hat h\left(\Phi(t_i),t_i+\frac{\Delta t}{2}\right),
\end{equation}
and applying the Crank-Nicolson method, the orbitals  at the next time step $t_{i+1}=t_i+\Delta t$ can be obtained by solving
\begin{equation}\label{eq:OrbitalCN}
\tilde A_{i}\varphi_j(t_{i+1})= \tilde B_{i}\varphi_j(t_{i}) \quad \text{for each orbital $\varphi_j$}.
\end{equation}
which implies that
\begin{equation}
    \norm{\tilde A_{i}\varphi_j(t_{i+1})- \tilde B_{i}\varphi_j\left(t_i\right)}^2=0 \quad \text{for each orbital $\varphi_j$},
\end{equation}
which in turn yields that the orbitals at the next time step fulfill 
\begin{equation}
    \sum_{j=1}^N\norm{\tilde A_{i}\varphi_j(t_{i+1})- \tilde B_{i}\varphi_j\left(t_i\right)}^2=0.
\end{equation}
Recasting the orbital time evolution equations as a minimization problem, the orbitals at the next time step $\Phi(t_{i+1})$ can be obtained as
\begin{equation}\label{eq:Abstract_Rothe_HF}
    \Phi(t_{i+1})=\argmin_{\Omega}\sum_{j=1}^N\norm{\tilde A_{i}\omega_j- \tilde B_{i}\varphi_j\left(t_i\right)}^2.
\end{equation}
where $\Omega=\{\omega_j\}_{j=1}^N$. 

We can now expand the orbitals $\omega_j$ in a shared one-particle basis $\left\{g_m(\bm{\alpha})\right\}_{m=1}^{M(t_{i+1})}$ as
\begin{equation}
  \omega_j=\sum_{m=1}^{M(t_{i+1})} c_{j,m}g_m(\alpha).
\end{equation}
Note that the Gaussian basis is shared across all orbitals, with only the linear coefficients $c_{j,m}$ differing between orbitals, leading to a more compact representation. In this expansion, the optimization is now performed with respect to parameters $\alpha$ and linear coefficients $c$ (where $c$ is the coefficient matrix), 
\begin{equation}\label{eq:Rothe_opt_HF}
    {\alpha}(t_{i+1}),{c}(t_{i+1})=\argmin_{{\alpha},{c}}\mathrm r^2_{i+1}({\alpha},{c}),
\end{equation}
where we defined the all-orbital Rothe error $r_{i+1}(\alpha,c)$
\begin{equation}\label{eq:RotheError_HF}
    \mathrm r^2_{i+1}({\alpha},{c})={\sum_{j=1}^N \tilde {\mathrm r}_{i+1}^2\left({\alpha},{c};\varphi_j\right)},
\end{equation}
which in turn consists of single-orbital Rothe errors $\tilde {\mathrm r}_{i+1}^2\left({\alpha},{c};\varphi_j\right)$, defined as 
\begin{equation}
\tilde {\mathrm r}_{i+1}( \alpha, c;\varphi_j)=\norm{\sum_{m=1}^{M(t_{i+1})} c_{j,m}\tilde A_{i}g_m({\alpha})- \tilde B_{i}\varphi_j\left(t_i\right)}.
\end{equation}
The all-orbital Rothe error is a measure of the overall error in the time evolution of the wave function, summing over the errors in each individual orbital. Conveniently, optimization over individual rows of the coefficient matrix $c_{j,{:}}$ for orbital $\varphi_j$ is independent of the other orbitals, as the all-electron Rothe error does not couple orbitals. Hence, Variable Projection can be used to obtain the whole coefficient matrix $c(\alpha)$ analytically, leaving an optimization problem with respect to $\alpha$ only.

The all-orbital Rothe error is no longer an explicit upper bound of the time-evolution error of the wave function as a whole, as coupling terms between orbitals are neglected. This is, however, not a problem in practice (see Sec. \ref{sec:TDHF_Rothe}), and the all-orbital Rothe error serves as a good proxy for the time evolution error and is a meaningful quantity to minimize to obtain correct dynamics, as has been established in Ref.~\cite{schraderTimeDependentGaussianBasis2025}. 

\subsection{Basis size flexibility}\label{sec:BasisFlexibility}

As pointed out in Eq. \eqref{eq:WF_ansatz}, the number of basis functions $M(t)$ does not need to be kept constant. Since the Rothe error \eqref{eq:RotheError} is an upper bound of the time evolution error, ensuring that the cumulative Rothe error at $t_n$ stays below a prescribed tolerance $\varepsilon$ guarantees that the wave function error at $t_n$ is at most $\varepsilon$. In order to control the time evolution error, this upper bound $\varepsilon$ can be made a parameter in Rothe's method. By the triangle inequality, the cumulative propagation error $t_n$ is bounded by the sum of the per-timestep Rothe errors,
$$
    \|\tilde{\Psi}(t_{n}) - \Psi(t_{n})\| \leq \sum_{i=1}^n \mathrm{r}_{i} \leq \varepsilon.
$$
Letting $N_{t_f}=t_f/\Delta t$ denote the total number of time steps, $\varepsilon_{\Delta t}=\varepsilon/N_{t_f}$ then serves as an upper bound on the Rothe error per time step. This parameter yields a natural way for the addition of basis functions: If the optimized Rothe error $r_i$ at time point $t_i$ is above $\varepsilon_{\Delta t}$, one can add a basis function and re-optimize the parameters of the expanded basis. Similarly, if removing a basis function still yields a Rothe error below $\varepsilon_{\Delta t}$, a basis function can be removed. A small value $\varepsilon$ requires more basis functions, but leads to presumably improved dynamics and results, while also increasing computational cost.

See Section~\ref{sec:addition and removal of gaussians} for a practical algorithm for adding and removing Gaussians.

\section{Implementation of Rothe's method with Gaussian basis functions}\label{sectionn: implementation}

\subsection{Required matrix elements}

For Rothe's method, matrix elements involving the identity $\hat{I}$, the Hamiltonian $\hat{H}(t)$ and the squared Hamiltonian $\hat{H}^2(t)$ are required, as described in Sec.~\ref{sec:VarPro}. Expressing the Hamiltonian as
\begin{equation}
    \hat{H}(t)=\hat{T}+\hat{V}(t),
\end{equation}
we see that the following matrix elements are required in order to assemble $\rho(\alpha)$ and $\check S(\alpha)$, see Sec.~\ref{sec:rothe for gaussians}: Let $\hat{X}$ be any of the operators in the set
$$
    \{ \hat{I}, \; \hat{T}, \; \hat{V}, \; \hat{T}^2, \; \hat{T}\hat{V}, \; \hat{V}\hat{T}, \; \hat{V}^2 \}.
$$
Let $\lambda$ and $\lambda'$ be given Gaussian parameter sets. For assembling the Rothe error, one needs algorithms for computing
$$
    \braket{\gamma(\lambda)|\hat{X}|\gamma(\lambda')}.
$$
For assembling the gradient of the Rothe error, one additionally needs the parameter derivatives
$$
    \frac{\partial}{\partial \lambda_\ell}     \braket{\gamma(\lambda)|\hat{X}|\gamma(\lambda')}.
$$
The latter can conveniently be obtained by differentiating the analytical expressions or by using libraries for automatic differentiation, as the expressions quickly become algebraically involved.

\subsection{Calculation of matrix elements}

We here describe an outline of how to obtain some of the needed gaussian matrix elements for the case of the potential being a sum of Coulomb terms, gaussians, and polynomials.

\paragraph{Gaussian overlap matrix elements.} For the matrix elements of $\hat{I}$, i.e., the overlap matrix of the Gaussian basis, we use the Gaussian product theorem: The product of two Gaussians is itself a Gaussian. For two Gaussians $g_n,g_m$, there exist unique $C_{nm}\in\mathbb C^{D\times D}$ and  $y_{nm}\in \mathbb C^D,d_{nm}\in \mathbb C$ 
such that
\begin{align}\label{eq:GaussianProductTheorem}
    g_n^*g_m=d_{nm}\exp\bigl(-x^{\top} C_{nm} x+ y_{nm}^\top x\bigr). 
\end{align}
The overlap between two $D$-dimensional Gaussians can be calculated as 
\begin{equation}\label{eq:OverlapFormula}
\begin{aligned}
    S_{nm}=\langle{g_n|g_m}\rangle&=\int d_{nm}\exp\bigl(-x^{\top} C_{nm} x + y_{nm}^\top x\bigr)\mathrm d x\\
    &= d_{nm} \frac{\pi^{D/2}}{\det\left(\sqrt{C_{nm}}\right)}\exp\left(\frac{1}{4} y_{nm}^\top C_{nm}^{-1} y_{nm}\right).
\end{aligned} 
\end{equation}
where $\sqrt{\cdot}$ is the principal matrix square root.

\paragraph{Polynomial terms.}
Matrix elements of polynomials can be calculated by repeated differentiation under the integral sign. Letting $x_k$ stand for the $k$'th element in $x$, one finds
\begin{equation}
\begin{aligned}
    \langle{g_n|x_k|g_m}\rangle&=\int d_{nm} x_k\exp\bigl(-x^{\top} C_{nm} x + y_{nm}^\top x\bigr)\mathrm d x\\
    &=\int  d_{nm}\frac{\partial}{\partial (y_{nm})_k}\exp\bigl(-x^{\top}  C_{nm} x + y_{nm}^\top x\bigr)\mathrm dx\\
    &=\frac{\partial}{\partial (y_{nm})_k}\int d_{nm}\exp\bigl(-x^{\top} C_{nm} x + y_{nm}^\top x\bigr)\mathrm dx \\
    &=\frac{\partial S_{nm}}{\partial (y_{nm})_k}.
\end{aligned}
\end{equation}
By repeatedly differentiating under the integral sign, this strategy can be used to calculate polynomial expectation values of arbitrary monomials.

\paragraph{Momentum and kinetic energy terms}
It holds that
$$
    \frac{\partial}{\partial x_i} \gamma(x;\lambda) = p_i(x;\lambda) \gamma(x;\lambda),
$$
where $p_i(x;\lambda)=\partial (\ln \gamma(x;\lambda))/\partial x_i$, a polynomial of degree at most 1. Thus any matrix element of any polynomial $Q(\partial/\partial x_1,\cdots,\partial/\partial x_D)$ of degree $d$ becomes a matrix element of a polynomial $P(x;\lambda)$ of same degree or lower.

In particular, for a kinetic energy operator in the form
\begin{equation}
    \hat{T}=-\sum_{i,j=1}^D \frac{1}{2m_{ij}}\frac{\partial^2}{\partial x_i \partial x_j },
\end{equation}
where $[m_{ij}]$ is a symmetric matrix, we get
$$
    \braket{\gamma(\lambda)|\hat{T}|\gamma(\lambda')} = \sum_{ij} \frac{1}{2m_{ij}} \braket{\gamma(\lambda)|p_i(\lambda)^* p_j(\lambda')|\gamma(\lambda')}.
$$

\paragraph{Gaussian potential terms.}
Many potentials are well approximated by linear combinations of Gaussians:
$$
    V(x) = \sum_{\nu=1}^{N_{\text{pot}}} w_\nu g_{\nu}(x),
$$
where the Gaussians $g_\nu$ and the linear coefficients $w_\nu$ are fixed and obtained by, e.g., polynomial fitting. By appealing to Eqs.~\eqref{eq:GaussianProductTheorem} and~\eqref{eq:OverlapFormula}, matrix elements of $V(x)$ between the Gaussian basis functions $g_n$ and $g_m$ are straightforwardly evaluated.

\paragraph{Polynomial times potential matrix elements.}

For any potential operator $V(x)$, where the matrix element $\braket{g_n| V(x)|g_m}$ is available, the matrix element of $P(x)V(x)$ where $P(x)$ is any polynomial is readily obtained by differentiation under the integral sign:
\begin{equation}
    \langle{g_n|x_k V(x)|g_m}\rangle=\frac{\partial }{\partial (y_{nm})_k}\langle{g_n| V(x)|g_m}\rangle,
\end{equation}
where $y_{nm}$ is defined in Eq.~\eqref{eq:GaussianProductTheorem}. This method can then be used to compute matrix elements of, say, $\hat{T}\hat{V}$.

\paragraph{Coulomb terms.}
For multiparticle systems, $x = (r_1,\cdots,r_N)$ with each $r_i \in \mathbb{R}^3$. The potential $\hat{V}(t)$ then includes Coulomb interactions. To keep these analytically solvable in the multiparticle case, one usually makes the choice
\begin{equation}\label{eq:CoulombCompatibleForm}
    {A}_m+\ii {B}_m= \left(A'_m + \ii B'_m\right)\otimes  I_3,
\end{equation}
where $I_3$ is the $3\times 3$ identity matrix and $\otimes$ is the Kronecker product.  For $3N$-dimensional Gaussians of the form of Eq.~\eqref{eq:CoulombCompatibleForm}, Coulomb integrals are analytically solvable. Let $r_i$ be the 3 consecutive components of $x$ belonging to particle $i$, and note the identity
\begin{equation}\label{eq:Coulomb_transform}
    {|{r_i- r_j}|}^{-1} = \frac{2}{\sqrt{\pi}} \int_0^{\infty} \exp\left({-s^2 \left({r_i- r_j}\right)^2}\right) \, \mathrm ds,
\end{equation}
as well as the fact that one can write
\begin{equation}
    |{ r_i- r_j}|^2= x^\top  J_{ij} x,
\end{equation}
\begin{equation}
    J_{ij}
    = \left((e_i-e_j)(e_i-e_j)^\top\right) \otimes I_3
\end{equation}
with $e_k$ being the $k$th standard basis vector of $\mathbb{R}^N$. This yields
\begin{equation}
\begin{aligned}
    &\langle{g_n|\left({\abs{ r_i- r_j}}\right)^{-1}|g_m}\rangle
    =\int \left({\abs{ r_i- r_j}}\right)^{-1} d_{nm}\exp\bigl(-x^{\top} C_{nm} x+  y_{nm}^\top x\bigr)\mathrm d x\\
    &\quad =\frac{2}{\sqrt{\pi}}\int \int_0^\infty d_{nm}\exp\bigl(-x^{\top}  C_{nm} x +  y_{nm}^\top x-s^2 \left({ r_i- r_j}\right)^2\bigr)ds \mathrm d x\\
    &\quad =\frac{2 d_{nm}}{\sqrt{\pi}}\int \int_0^\infty\exp\bigl(-{x}^{\top} \left( C_{nm}+s^2{J}_{ij}\right)x+ y_{nm}^\top  x\bigr)\mathrm ds \mathrm d x.
\end{aligned}
\end{equation}
The $3N$-dimensional integral over $x$ can be solved using the overlap integral formula, giving rise to a one-dimensional integral in $s$, which is analytically solvable \cite{Bubin_Adamowicz_ECG2013,Kinghorn1995}.

Matrix elements involving products of polynomial and Coulomb terms such as $x_k\left({\abs{ r_i- r_j}}\right)^{-1}$, which arise in terms such as $\hat{V}(t)\hat T$ and $\hat{V}^2(t)$, can also be implemented by differentiating under the integral sign, see above.

Finally, we consider squared Coulomb operators like
$\left({\abs{ r_i- r_j}}\right)^{-1}\left({\abs{ r_k- r_l}}\right)^{-1}$ which arise in $\hat{V}^2(t)$. To our knowledge, analytical expressions for expectation values involving ECGs only exist for EGCs that share a center \cite{ireland2021integralslowerboundsexact}. However, the integral can be well approximated by expressing one of the Coulomb terms as a linear combination of $N_g$ Gaussians, i.e.,
\begin{equation}\label{eq:Coulomb_as_Gaussians}
\left({\abs{r_i- r_j}}\right)^{-1}\approx\sum_{k=1}^{N_g}w_ke^{-p_k\abs{ r_i- r_j}^2}.
\end{equation}
There are efficient algorithms to evaluate the parameters $w_k$ and $p_k$, and rapid convergence with increasing $N_g$ is possible \cite{beylkin2005approximation}. \citeauthor{racsai2024regularized} have used this approximation for calculations using ECGs in molecules \cite{racsai2024regularized}. With a potential of the form of Eq. \eqref{eq:Coulomb_as_Gaussians}, integrals can be evaluated analytically, as the integrand will be a Gaussian due to the Gaussian product rule, i.e.,  
\begin{equation}
    g_n^*\left(w_ke^{-p_k\abs{ r_i- r_j}^2}\right)g_m= d_{nkm}\exp\bigl(-{x}^{\top} {C}_{nkm}{x}+ y_{nkm}^\top  x\bigr).
\end{equation}
This integral can be evaluated using Eq. \eqref{eq:OverlapFormula}.

\subsection{Numerical considerations}

To prevent the linear coefficients $\bm c$ from spanning several orders of magnitude, which can lead to instability, one should work with normalized Gaussians.

The squared Rothe error $r_i^2$ arises naturally from the derivation and serves as the objective function to be minimized. With a total of $10\,000$ time steps, an average Rothe error of $10^{-4}$ corresponds to a cumulative Rothe error of $1$ at $t=T_{f}$. The worst-case bound on the wave function error is then of order unity, providing no useful guarantee on the quality of the approximation. In practice, we have observed that even double-digit final Rothe errors still yield quantitatively correct spectra and densities, indicating that the Rothe error can put substantial weight on (local) wavefunction errors that may not be decisive for the quantities of interest in a given simulation. This is well known for time-independent Hamiltonians in electronic-structure theory where the energy may be very well converged while the energy variance is much greater than zero. For time-dependent Hamiltonians, on the other hand, the variance should not vanish and the Rothe error may turn out to be an overly conservative upper bound to the wavefunction error.
\paragraph{Scale-invariance and Hessian initialization.}
The Rothe error can vary over orders of magnitude as part of an optimization trajectory, in the sense that $\mathrm r_{i+1}( \alpha_{\rm init})\gg\mathrm r_{i+1}( \alpha_{\rm final})$, where $\alpha_{\rm init}$ is the start guess and $\alpha_{\rm final}$ the converged parameters. Hence, a gradient-based optimization algorithm should be scale-invariant in order for the update
steps not to depend on the magnitude of the Rothe error or its gradient.
The Broyden–Fletcher–Goldfarb–Shanno (BFGS) algorithm and its limited memory variant are considered to be among the best quasi-Newton optimization algorithms for non-stochastic optimization problems with thousands of parameters \cite{nocedal_numerical}. The BFGS algorithm is scale-invariant when optimal line search is performed~\cite{powellConvergenceVariableMetric1971}.

However, this is often not the case for numerical implementations, and the scale-invariance breaks. Hence, the initial Hessian inverse $H_0$ needs to be initialized carefully. Since the squared Rothe error and its gradient are typically very small, setting $H_0=I$ leads to a tiny initial step $p_0 = - H_0 \nabla_{\alpha}\mathrm{r}^2_{i+1}\left(\alpha^{\rm init}_{i}\right)$ and slow convergence. Setting instead
\begin{equation}
     H_0=\|{\nabla_{\alpha}\mathrm{r}^2_{i+1}\left(\alpha^{\rm init}_{i}\right)}\|^{-1} I,
\end{equation}
makes the initial step invariant under rescaling of the Rothe error by a constant $c$, restoring a sensible step length.
\paragraph{Initial parameter guess.} 
Another question is how to correctly initialize the parameters $\alpha^{\rm init}_{i}$. For small time steps $\Delta t$, the solution from the previous time step, ${\alpha}(t_{i-1})$, is often a reasonable start guess. Assuming that changes in the parameters are approximately constant between time steps, i.e, ${\alpha}(t_{i})-{\alpha}(t_{i-1})\approx {\alpha}(t_{i-1})-{\alpha}(t_{i-2})$, an improved start guess is 
$\alpha^{\rm init}_{i}(p)=(1+p) \alpha(t_{i-1})-p  \alpha(t_{i-2})$, where $p$ is found by line search.
\paragraph{Parameter bounds.}
For stability, the parameters should not change too much between time steps. In particular, if the optimization procedure changes parameters too aggressively, some Gaussians might end up in regions where they no longer contribute to the wave function. One can restrict this behavior by explicitly enforcing bounds on the parameters, e.g., by carrying out optimization on a transformed set of parameters, such as \cite{schraderTimeEvolutionOptimization2024,schraderMultidimensionalQuantumDynamics2025}
\begin{align}
    ({\alpha}_{i+1})_j  = ({\alpha}^{\min}_{i+1})_j
    &+ \big(\tanh( ({\alpha}_{i+1})'_j)\! +\! 1\big) \frac{\big(({\alpha}^{\max}_{i+1})_j - ({\alpha}^{\min}_{i+1})_j\big)}{2}.
 \end{align}
Optimization is then carried out with respect to the transformed parameters ${\alpha}_{i+1}'$, and parameters are guaranteed to remain in the range $\left[{\alpha}^{\min}_{i+1},{\alpha}^{\max}_{i+1}\right]$.

\paragraph{Frozen parameters for eigenstates.} Furthermore, when starting a calculation from an eigenstate of the Hamiltonian, it can be helpful to \emph{freeze} the parameters that represent the Gaussians of that state. The reason is that even small perturbations in these parameters can lead to large changes in the variance and hence the Rothe error, destabilizing the optimization procedure.

\paragraph{Near-linear dependence.} 
The perhaps most severe numerical stability issue, however, is near-linear dependency between weighted Gaussians, leading to very small eigenvalues in $\check { S}= S+O(\Delta t^2)$. This problem can be partially mitigated using Tikhonov regularization \cite{tikhonov_solutions_1977,doicu_numerical_2010_chapter3}, i.e., obtaining the linear coefficients as 
\begin{equation}
     c( \alpha)=\left(\check { S}( \alpha)+\lambda  I\right)^{-1} \rho( \alpha), \label{eq:tikhonov}
\end{equation}
where $\lambda$ is a small parameter,chosen in the range $10^{-6}$ to $10^{-10}$. Alternatively, one can obtain a new set of Gaussians that are less linearly dependent by refitting a new set of Gaussians to the wave function. This should be done before Rothe optimization and the Rothe error should be tracked carefully---if it is of the same order of magnitude as the Rothe error at previous time steps, the error due to refitting is negligible. 

We stress that the regularization~\eqref{eq:tikhonov} is only a trick to overcome near linear dependence during optimization, and does \emph{not} affect the basic equation, i.e., minimization of the rothe error. The effect of the regularization is to choose slightly suboptimal linear coefficients -- the wavefunction $\Psi(\alpha)$ is still well-defined with a well-defined Rothe error. This should be contrasted with the similar regularization defined in Eq. \eqref{eq:regularization hack in MVP} for the MVP, which changes the equation being solved, and hence affects accuracy.

Linear dependency can arise between pairs of Gaussians, i.e., $\abs{\braket{g_n|g_m}}\approx1$ ($n\neq m$). Once two Gaussians are almost identical, the gradients of the Rothe error with respect to each Gaussian's parameters will also be near-identical and the Gaussians will remain close. To avoid this issue, Gaussians can be refitted, or a penalty term can be added to the optimization, such that there is an extra term added to the optimization function when $\abs{\braket{g_n|g_m}}>s$, where $s$ can be chosen to be $\sim 0.99$.

\subsection{Addition and removal of Gaussians.}
\label{sec:addition and removal of gaussians}

As pointed out in Sec. \ref{sec:BasisFlexibility}, Rothe's method naturally allows for a variable number of basis functions. In particular, if, after optimization, the Rothe error is greater than the threshold $\varepsilon_{\Delta t}$ at a time step $t_{i+1}$, a Gaussian can be added. To suggest new Gaussian parameters, one can use the Stochastic Variational Method \cite{Bubin_Adamowicz_ECG2013}, where candidate Gaussians are drawn from a distribution
\begin{equation}
p({\alpha}^{(k)}_{M(t_{i})+1})\propto\sum_{m=1}^{M(t_{i})}\exp\!\left[-\frac{({\alpha}^{(k)}_{M(t_{i})+1}-{\alpha}_{m})^2}{2{\alpha}_m^2}\right].
\end{equation}
This expression should be understood element-wise. Here, $\alpha_m$ are the parameters that already are part of the basis. One then selects the Gaussian $g({\alpha}_{M(t_i)+1}^{(k^{\mathrm{opt}})})$ that gives rise to the lowest Rothe error, and adds it to the basis. After this, all parameters $\alpha$ are re-optimized. 

To speed up Rothe calculations and to avoid linear dependency, one can remove Gaussians that do not meaningfully contribute to the Rothe error, i.e., Gaussians that, if removed, barely increase the Rothe error.

\subsection{Norm  and energy conservation}

While the Crank-Nicolson method is norm- and energy-conserving, Rothe's method merely approximates the Crank-Nicolson solution when the basis is not complete. Consequently, it is not inherently norm- and energy-conserving. However, conservation can be enforced explicitly in Rothe's method. One can carry out a constrained optimization subject to norm  and energy conservation using Lagrange multipliers, albeit at the cost of making VarPro unusable. A simple workaround is to first carry out an unconstrained Rothe optimization to obtain the optimal nonlinear parameters $\alpha(t_{i+1})$, and then re-determine the linear coefficients $c(t_{i+1})$ at fixed $\alpha(t_{i+1})$ by a constrained minimization, subject to conservation of norm and energy (for time-in\-dep\-en\-dent Hamil\-tonians), i.e.,
\begin{equation}\label{eq:reorthonormalize}
\begin{aligned}
    &{c}(t_{i+1}) = \argmin_{{c}} \;  \mathrm r^2_{i+1}({\alpha}(t_{i+1}), {c}),  \\
    &\text{subject to}\ \left\{
    \begin{aligned}
        &\langle\Psi(t_{i+1})|\Psi(t_{i+1})\rangle - \langle\Psi(t_{i})|\Psi(t_{i})\rangle = 0, \\
        &\langle\Psi(t_{i+1})|\hat{H}|\Psi(t_{i+1})\rangle - \langle\Psi(t_{i})|\hat{H}|\Psi(t_{i})\rangle = 0.
    \end{aligned}
    \right.
\end{aligned}
\end{equation}
For norm conservation only, the simplest approach is to rescale the linear coefficients obtained using VarPro. 

For TDDFT and TDHF theory, one can make sure that orbitals remain orthonormal by using Löwdin symmetric orthogonalization \cite{Mayer_Lowdin} for orthogonalization followed by rescaling of the coefficient matrix $\bm c(t_{i+1})$ for normalization. 

\subsection{Masks} 

\begin{figure}
    \centering
    \includegraphics[width=0.5\linewidth]{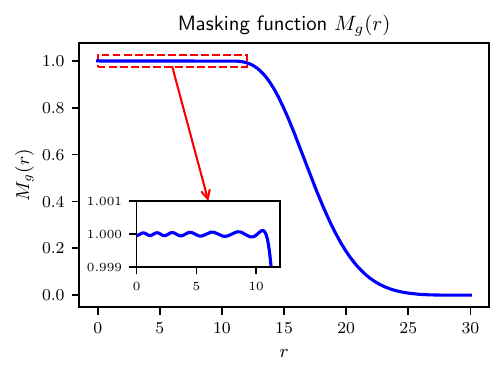}
    \caption{An example of a Gaussian mask consisting of $20$ Gaussians.
    (Reprinted from Ref.~\cite{schraderMultidimensionalQuantumDynamics2025} with the permission of AIP Publishing.)}
    \label{fig:GaussMask}
\end{figure}

To reduce computational cost, but also as a method to calculate ionization-dependent properties, one can use \emph{masks} to absorb an outgoing wave function~\cite{kosloffAbsorbingBoundariesWave1986a,neuhasuerTimedependentSchrodingerEquation1989}. A mask $\mathrm{Mask}(x)$ is a function that is equal to $1$ in a region of interest, $0$ in a region not of interest, and smoothly decays in the intermediate region. The effect of a mask is that the wave function, after each time step, is replaced by its masked version, i.e., 
\begin{equation}\label{eq:Mask}
    \Psi(x,t) \leftarrow \mathrm{Mask}(x)\Psi(x,t).
\end{equation}
Since $\Psi(x,t)$ is a linear combination of Gaussians, expressing $\mathrm{Mask}(x)$ as a linear combination of Gaussians ensures that the masked wave function remains a linear combination of Gaussians, which can then be analytically obtained. To obtain an expression for $\mathrm{Mask}(x)$, one can pick a mask, such as a standard trigonometric $\cos^{1/8}$-type mask, and approximate it as a linear combination of Gaussians:
\begin{equation}
    \mathrm{Mask}(x)\approx \mathrm{Mask}_g(x)=\sum_{m=1}^{N_{mg}}w_me^{-(x-\mu_m)^\top A_m(x-\mu_m)},
\end{equation}
where the real parameters $\mu_m,A_m,w_m$ can be obtained using curve fitting, minimizing the residual  $\|\mathrm{Mask}-\mathrm{Mask}_g\|^2$. After masking, the wave function consists of $M(t_i)\cdot N_{mg}$ Gaussians. To prevent unbounded growth in the basis size, this expanded representation is re-fitted to a linear combination of $M(t_i)$ Gaussians at each time step. An example of a mask expressed as a linear combination of Gaussians is illustrated in Fig. \ref{fig:GaussMask}.
\subsection{Computational considerations}
\paragraph{Computational cost.} The computational cost of Rothe's method depends on the choice of basis functions (for example, whether Gaussians are correlated or not), the potential, and the dimensionality $D$ of the problem. In general, the term that dominates the cost is the repeated evaluation of the matrix elements of the squared Hamiltonian $\hat{H}^2(t)$ and their derivatives due to their complexity. For $M$ basis functions with $K$ parameters per Gaussian ($K$ scales as $O(D^2)$ for ECGs and as $O(D)$ for uncorrelated Gaussians), the computational cost scales asymptotically as $O(M^3)$ due to the linear solve to obtain the linear coefficients. However, unless $M$ is very large, the computational cost is practically dominated by the evaluation of $O(M^2)$ matrix elements and a total of $O(M^2K)$ derivatives. To obtain sufficient convergence of the optimization problem, one needs a lot of iterations, and we found that approximately $300$ iterations per time step are necessary, to obtain sufficiently converged Rothe errors when using the BFGS algorithm \cite{schraderTimeEvolutionOptimization2024,schraderMultidimensionalQuantumDynamics2025,schraderTimeDependentGaussianBasis2025}. The total parameter count is $N_p = O(MK)$. The BFGS algorithm has a memory footprint of $O(N_p^2)$, which can become a bottleneck once $N_p \sim 1000$, in which case one should switch to limited-memory BFGS \cite{LBFGS}.
\paragraph{Parallelizability.} We have implemented a parallel version of Rothe's method in Ref. \cite{schraderMultidimensionalQuantumDynamics2025}. The most time-consuming part, the evaluation of $O(M^2)$ matrix elements and their derivatives, can be carried out independently of other matrix elements, making the a major part of Rothe's method embarassingly parallel. With $C$ cores, each core calculates $O(M^2/C)$ matrix elements and their derivatives. All matrix elements are then gathered on a root core and the relevant matrices are assembled. Calculation of the Rothe error, its gradient, and the linear coefficients is then carried out on the root core, and a single optimization step is carried out. The updated coefficients and parameters are then broadcast to the worker cores and the process is repeated. 
\subsection{Coulomb potentials}\label{sec:Coulomb_variance}
\begin{figure}
    \centering
    \includegraphics[width=0.7\linewidth]{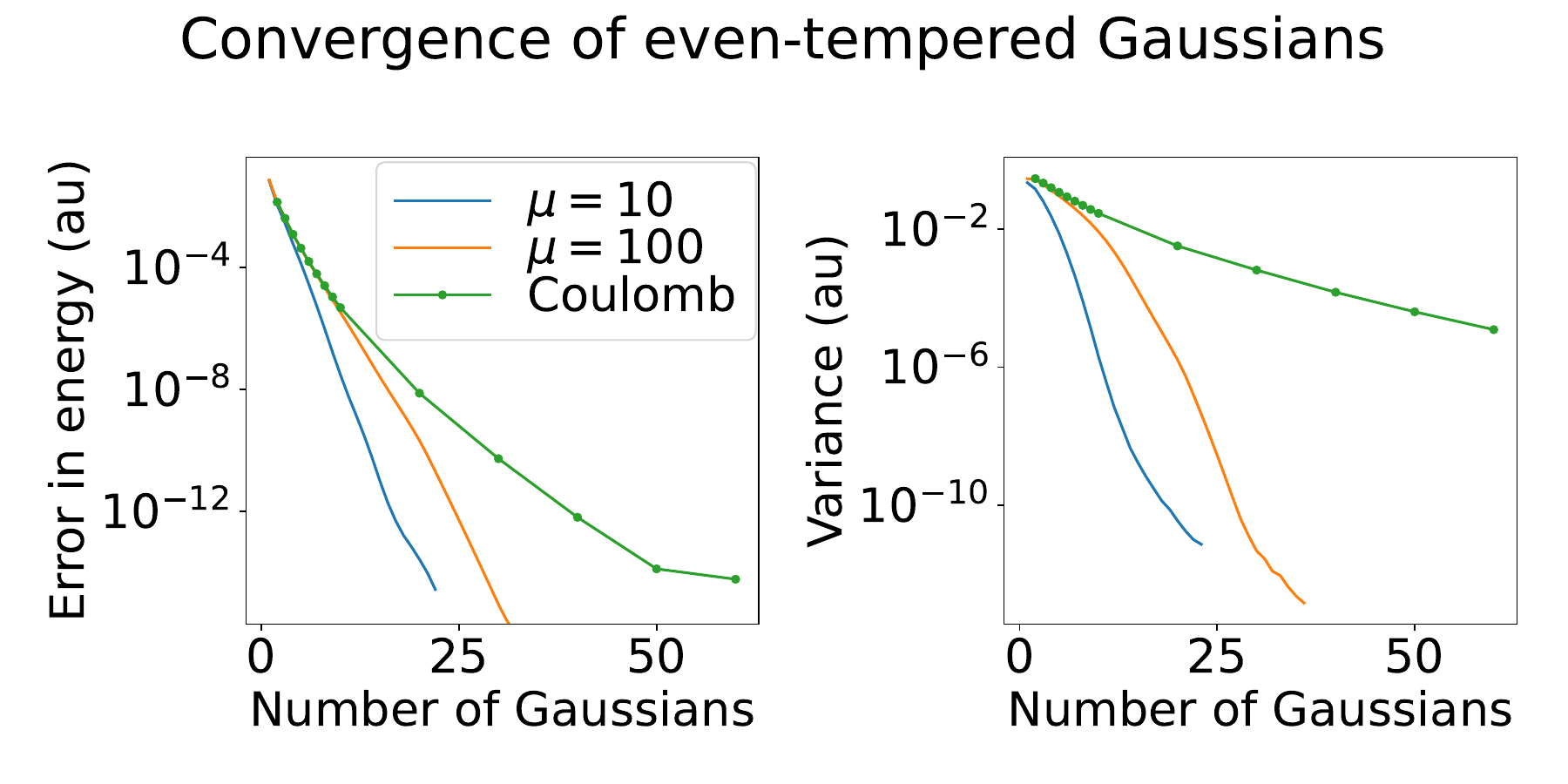}
    \caption{Singularity-free approximate Coulomb potentials allow for a faster convergence of the variance. Error in energy compared to exact ground state energy (left) and variance (right) as a function of the number of even-tempered Gaussians for the regularized (Eq. \eqref{eq:erf_potential}) and unregularized hydrogen atom Hamiltonians. 
    (Reprinted from Ref.~\cite{schraderTimeEvolutionOptimization2024} with the permission of AIP Publishing.)}
    \label{fig:variance}
\end{figure}
As pointed out in Sec. \ref{sec:Rothe_variance}, the Rothe error in time-independent potentials is proportional to the variance of the Hamiltonian. This is problematic for Coulomb potentials, since the corresponding wave functions exhibit cusps at coalescence points \cite{Kato_cusp}. The correct behavior cannot be represented using a finite number of Gaussians, and the variance will remain nonzero. Figure \ref{fig:variance} shows the convergence of the energy and the variance of the ground state as a function of the number of even-tempered Gaussians for the regularized hydrogen atom Hamiltonian
\begin{equation}\label{eq:erf_potential}
    \hat{H}(\mu)=-\frac{1}{2}\nabla^2-\frac{\erf (\mu r)}{r}.
\end{equation}
This Hamiltonian converges to the hydrogen atom Hamiltonian as $\mu \rightarrow \infty$. We observe that the variance converges very slowly for the unregularized Hamiltonian, yielding large Rothe errors. These large Rothe errors pose an issue, since the variance contribution dominates the dynamics contribution, and optimization then improves the ground-state representation rather than tracking the time evolution. However, we also see that the large variance can be mitigated by regularization of the Hamiltonian, with variances below $10^{-11}$ attainable with $25$ Gaussians for $\mu=100$.
Hence, applying Rothe's method to atomic and molecular systems requires using regularized Hamiltonians. In Ref. \cite{schraderTimeEvolutionOptimization2024}, it is shown that this regularization, with $\mu=100$, has a negligible effect on the HHG spectra for the hydrogen atom.

\subsection{Quadrature-based proof-of-concept implementation}
\label{sec:quadrature}

Rothe's method requires many matrix elements of operators between gaussian functions, the programming of which is error prone. Assembly of the Rothe error and its gradient is therefore costly. An alternative approach viable in low spatial dimensions ($D=1,2$) is to replace all inner products with quadrature. There are no longer any analytical integrals to be computed, and given that Gaussians are analytic functions, trapezoidal quadrature gives very high accuracy. We here outline this approach, which is useful for testing the Rothe method without having a fully optimized implementation.

The approach starts by introducing a rectangular uniform grid $G \subset [-L/2,L/2]^D \subset \mathbb{R}^D$ with $N_G^D$ points and volume element $\tau = [L/(N_G-1)]^D$. The continuous Hilbert space $L^2(\mathbb{R}^D)$ is then replaced by a finite-dimensional approximation, with inner product
$$
    \braket{\Phi|\Psi} \approx \tau \sum_{x\in G} \Phi(x)^* \Psi(x).
$$
The action of the projection operator $\hat{P}(\alpha)$ defined in Eq.~\eqref{eq:weighted Projection operator} and the evaluation of the Rothe error on the form of Eq.~\eqref{eq:Rothe error abstract form} can be efficiently computed using quadrature, as the need for computing matrix elements of $\hat{H}(t_{i+1/2})^2$ is eliminated. Instead, one directly computes inner products of the type $\braket{\hat{A}_i g_n|\hat{A}_i g_m}$ using the quadrature formula.

The quadrature based approach is useful for proof-of-concept studies, but beyond $D=2$ dimensions it becomes very time-consuming and thus ill-suited for judging computational efficiency relative to, e.g., grid-based methods. The quadrature-based approach was used in the applications described in Sec.~\ref{sec:applications}.

\section{Applications}
\label{sec:applications}

In this section, we present simulations of simplified model systems that validate key features of Rothe propagation of linear combinations of GWPs.
Our main focus is on electric-field-driven dynamics involving highly excited and/or continuum states, for electron dynamics and rovibrational
dynamics separately. We do not use optimized software for the evaluation of matrix elements (typically, we use quadrature as discussed in Sec.~\ref{sec:quadrature})
and, therefore, we do not focus on computational performance at this development stage.

\subsection{Strong-field electron dynamics}

The semiclassical Hamiltonian describing an atom or molecule interacting with an external electromagnetic field reads $\hat{H}(t)=\hat{H}_0+\hat{U}(t)$, where $\hat H_0$ is the field-free system Hamiltonian and $\hat{U}(t)$ describes the interaction with the field. Within the length-gauge electric-dipole approximation \cite{joachain_book}, $\hat{U}(t)$ reads
\begin{equation}
\hat{U}(t)= -\bm E(t) \cdot \hat{\bm D},
\end{equation}
where $\bm E(t)$ is the electric field and $\hat{\bm D}$ is the electric-dipole operator
\begin{equation}\label{eq:Dipole}
    \hat {\bm D}=\sum _{i=1}^Nq_i{\bm r}_i,
\end{equation}
where the sum goes over all $N$ particles with charge $q_i$.

To quantify the quality of the propagation obtained with Rothe's method, we compare with results obtained with highly accurate grid-based methods.
Expectation values are compared and, when reasonable, we compare the Rothe wavefunction with the grid one. Finally, we compare HHG spectra
calculated from the time-dependent electric-dipole moment as
\begin{equation}\label{eq:HHG_calc}
S(\omega) \propto \sum_{j\in\{x,y,z\}}\omega^2\left| \int_{0}^{T_f}  \,
\left(\langle \Psi(t) |\hat {\bm D}_j | \Psi(t) \rangle\right) \, e^{\ii \omega t}w(t)\mathrm{d}t \right|^2 .
\end{equation}
Here, $\bm E(t)$ is nonzero only for $0\leq t\leq T_f$ and $w(t)$ is a window function. 

\subsubsection{1D hydrogen atom}

Reference \cite{kvaal2023need} describes an initial application of Rothe's method with GWPs to a 1D model of the hydrogen atom, using the soft Coulomb potential
$V(x) = -(1/2)/\sqrt{x^2 + 1/4}$. The initial state is a linear combination of four GWPs centered at the origin with zero momentum and purely real covariance.
The covariances are obtained along with the linear expansion coefficients by least-squares fitting to a highly accurate ground-state wavefunction computed by imaginary-time
propagation on a grid.
The resulting initial state in GWP basis is accurate to around $7$ digits (compared with the grid wavefunction). The atom is then exposed to a very strong electric
field of the form 
\begin{equation}
    \label{eq:electric field 1D}
    \bm E(t)=E_0\,f(t)\cos\!\left(\omega (t-\bar{t})\right), \qquad \bar{t} = \frac{t_0 + t_1}{2},
\end{equation}
with $\omega = 0.25\,\text{a.u.}$ and $E_0 = 0.225\,\text{a.u.}$.
The envelope $f(t) = \sin^2(\pi(t - t_0)/(t_1-t_0))$ is nonzero only
in the interval $t \in [t_0,t_1] = [20, 80]\,\text{a.u.}$. With time steps of $\Delta t = 10^{-3}\,\text{a.u.}$, 
the Rothe propagation starts at $t=0\,\text{a.u.}$ and ends at $t=100\,\text{a.u.}$.

\begin{figure}[htbp]
    \centering
    \includegraphics[width=0.48\textwidth]{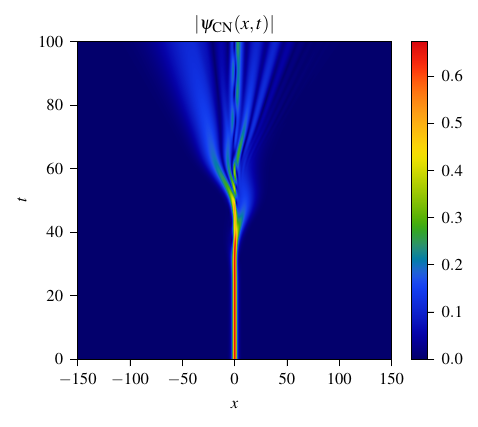}
    \hfill
    \includegraphics[width=0.48\textwidth]{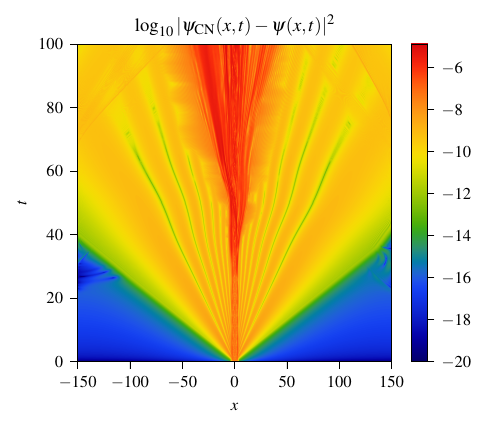}
    \caption{Left panel: Propagation history of the reference wavefunction obtained using the Crank-Nicolson integrator for the 1D hydrogen atom.
    Right panel: Absolute error of the wavefunction obtained by Rothe propagation.
    (Adapted from Ref. \cite{kvaal2023need}. Copyright  \copyright 2023 S. Kvaal, C. Lasser, T. B. Pedersen, and L. Adamowicz. Published by arXiv.org.)
    }
    \label{fig:1D_H_history}
\end{figure}
The wavefunction history obtained on a large grid is depicted in the left panel of Fig.~\ref{fig:1D_H_history} and the absolute error of the Rothe propagation is shown in the 
right panel. The wavefunction is initially localized around the origin, spreading over more than $100\,\text{a}_0$ as the system interacts with the laser pulse. After the
laser pulse has been switched off (at $t=80\,\text{a.u.}$), the wavefunction continues to spread, respresenting the amplitudes of the ionized electron. As is evident from the
right panel of Fig.~\ref{fig:1D_H_history}, the complicated wavefunction can be very accurately described by a linear combination of GWPs propagated by Rothe's method.
\begin{figure}[htbp]
    \centering
    \includegraphics[width=\textwidth]{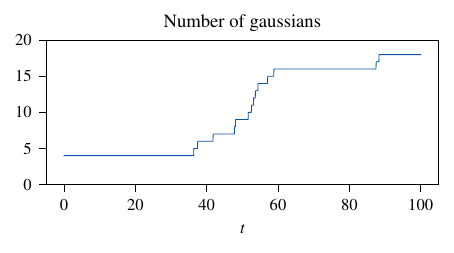}
    \caption{The number of GWPs required for Rothe optimization of the 1D hydrogen wavefunction.
    (Reprinted from Ref. \cite{kvaal2023need}. Copyright  \copyright 2023 S. Kvaal, C. Lasser, T. B. Pedersen, and L. Adamowicz. Published by arXiv.org.)
    }
    \label{fig:1D_H_N}
\end{figure}
Starting with $4$ GWPs for the initial ground-state wavefunction, Rothe's method adaptively adjusts the basis and a total of just $18$ GWPs are required for the final state,
see Fig.~\ref{fig:1D_H_N}. Note that GWPs are added also after interaction with the laser to capture the continued spreading of the wavefunction.

\subsubsection{2D hydrogen atom}

Going from 1D to 2D, not only radial but also rotational dynamics must be resolved, and to test the performance of GWP expansions with Rothe's method, Ref.~\cite{wozniak2025rothetimepropagationcoupled} considers a 2D hydrogen atom with the soft Coulomb potential $V(x,y) = -1/\sqrt{x^2 + y^2 + 1/2}$. The initial state is the 1s-like ground state,
determined variationally as a linear combination of $6$ GWPs centered at the origin with zero momentum and real covariance matrix proportional to the unit matrix. An $x$-polarized electric field of the form \eqref{eq:electric field 1D} with $\omega = 0.25\,\text{a.u.}$, and $E_0 = 0.4\,\text{a.u.}$ is applied from $t_0 = 0\,\text{a.u.}$ to $t_1 = 60\,\text{a.u.}$, and the wavefunction is propagated using Rothe's method with time step $0.002\,\text{a.u.}$ and with three different convergence thresholds for the minimization problem in each time step, $10^{-3}$, $10^{-4}$, and $10^{-5}\,\text{a.u.}$.
An analogous simulation is carried out with a grid-based wavefunction (using the Crank-Nicolson integrator) for reference.

\begin{figure}[htbp]
    \centering
    \includegraphics[width=\textwidth]{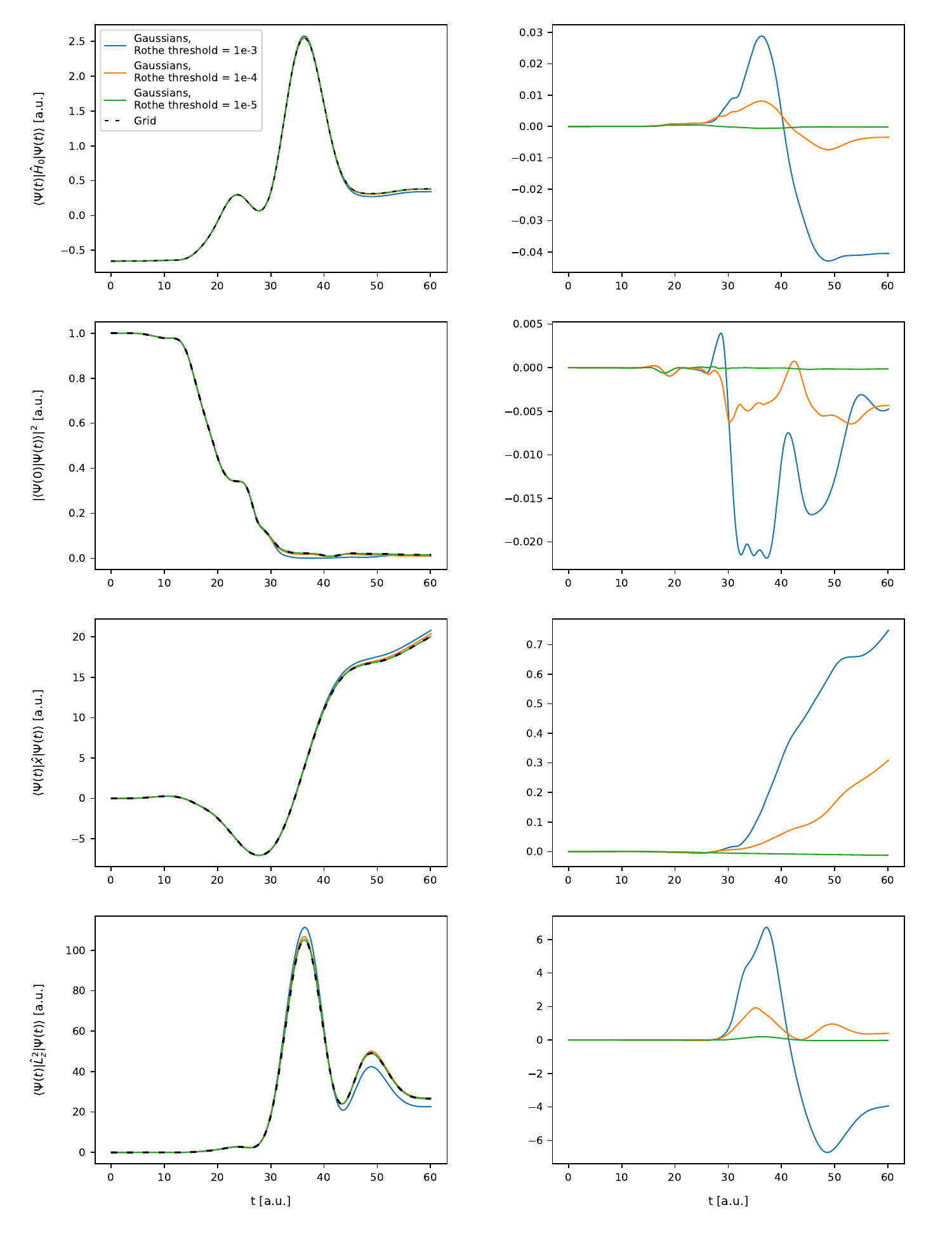}
    \caption{Left column: expectation values as functions of time for three values the Rothe minimization threshold along with the grid-based reference results for the
        2D hydrogen atom.
        Right column: Errors with respect to the reference expectation values. Rows from top to bottom contain results for the energy, ground-state survival probability,
        average position on the $x$-axis, and squared angular momentum, respectively.
        (Reprinted from Ref. \cite{wozniak2025rothetimepropagationcoupled}. Copyright  \copyright 2025 A. P. Wo{\'z}niak, L. Adamowicz, T. B. Pedersen, and S. Kvaal. Published by American Chemical Society.)
        }
    \label{fig:2D_H_observables}
\end{figure}
Time-dependent observables are plotted in Fig.~\ref{fig:2D_H_observables}. For all three Rothe thresholds we observe a reasonable agreement with the grid reference, although
high accuracy throughout the propagation is only observed for the lowest threshold. Errors generally increase after the laser pulse has reached its maximum intensity at
$t = 30\,\text{a.u.}$, which is roughly the time at which the wavefunction becomes orthogonal to the ground-state. At $t>30\,\text{a.u.}$, the positive energy indicates an unbound state consistent with the increasing average ($x$) position beyond the bound region of the soft Coulomb potential. Although the GWPs are not angular-momentum eigenfunctions,
they manage to capture the angular momentum correctly due to the plane-wave factors as long as the Rothe threshold is chosen low enough.

\begin{figure}[htbp]
    \centering
    \includegraphics[width=\textwidth]{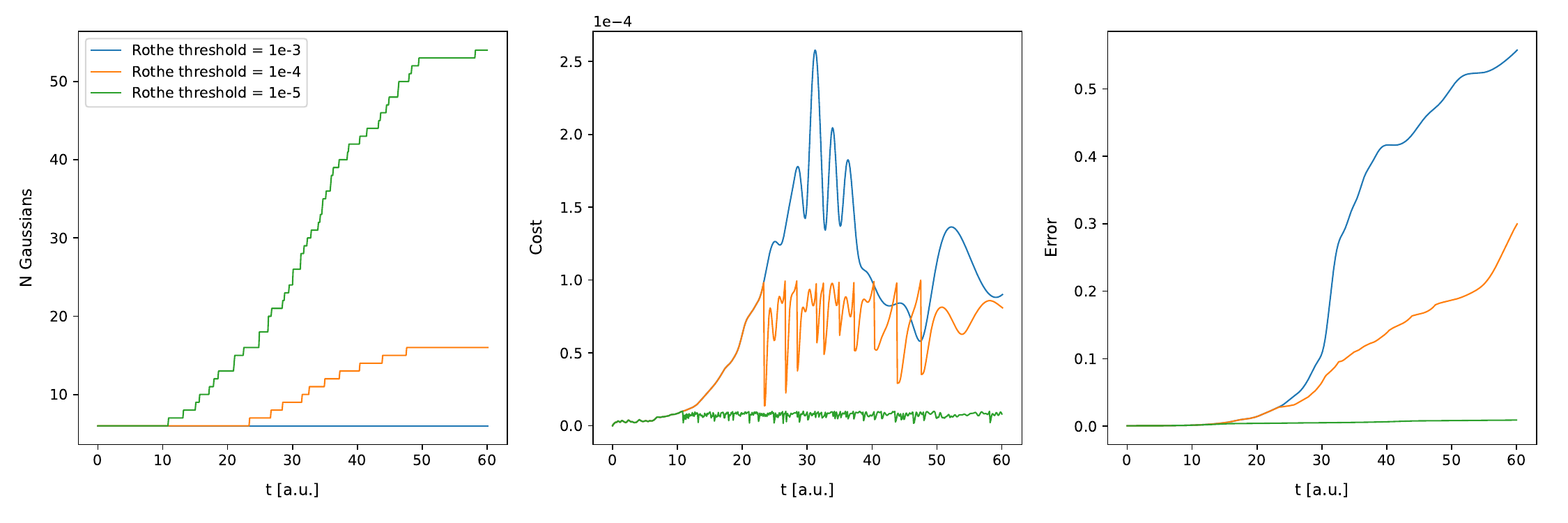}
    \caption{Left panel: Number of Gaussians required to achieve a certain level of accuracy for the 2D hydrogen atom. Middle panel: Value of the Rothe error in each time
        step. Right panel: Wavefunction error relative to the grid reference in each time step.
        (Reprinted from Ref. \cite{wozniak2025rothetimepropagationcoupled}. Copyright  \copyright 2025 A. P. Wo{\'z}niak, L. Adamowicz, T. B. Pedersen, and S. Kvaal. Published by American Chemical Society.)
        }
    \label{fig:2D_H_cost}
\end{figure}
\begin{figure}[htbp]
    \centering
    \includegraphics[width=\textwidth]{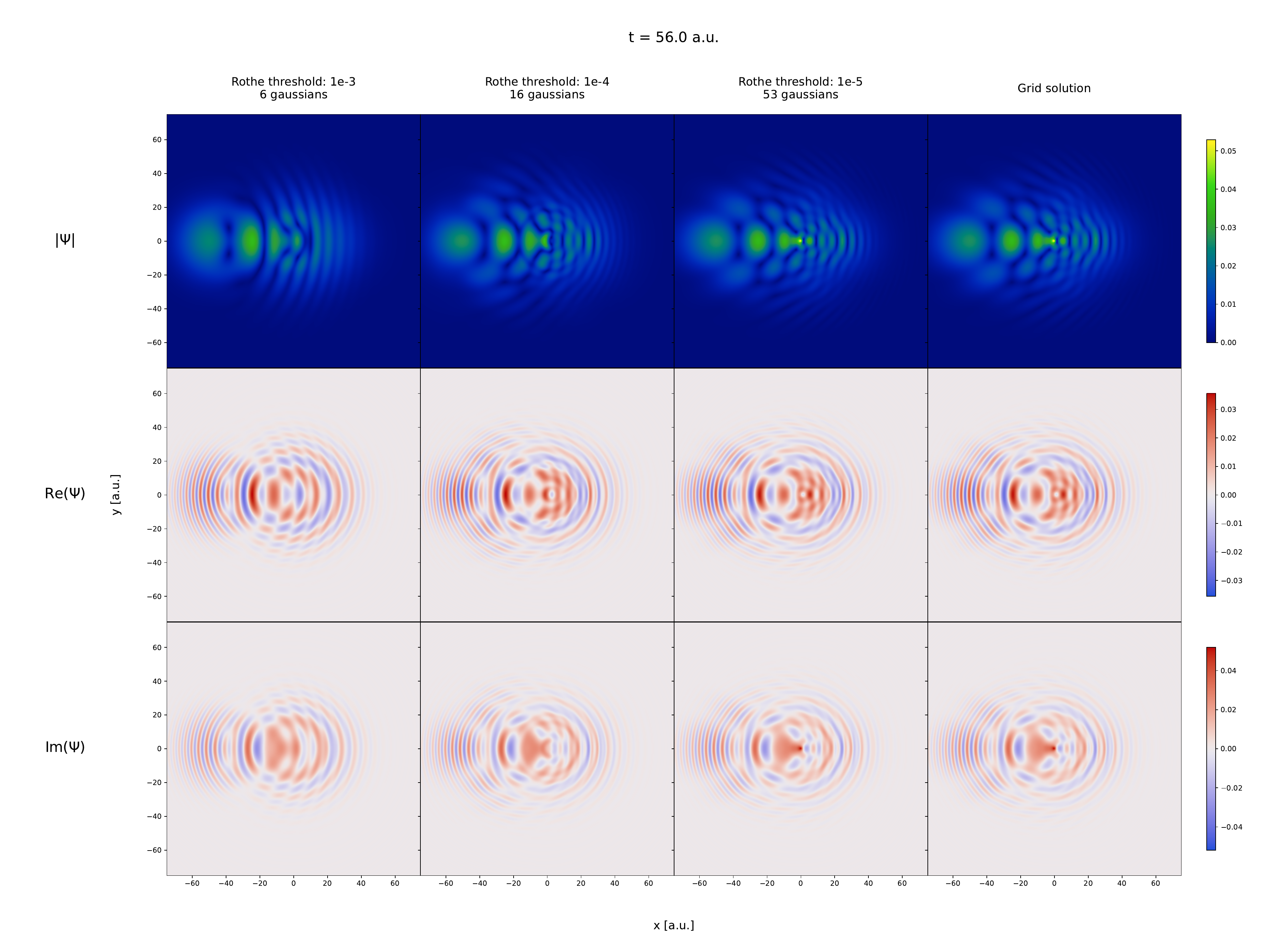}
    \caption{Plot of the absolute value and the real and imaginary parts of the 2D hydrogen wavefunction near the end of the laser pulse at $t=56\,\text{a.u.}$ for three different Rothe thresholds.
        The last column shows the reference wavefunction.
        (Reprinted from Ref. \cite{wozniak2025rothetimepropagationcoupled}. Copyright  \copyright 2025 A. P. Wo{\'z}niak, L. Adamowicz, T. B. Pedersen, and S. Kvaal. Published by American Chemical Society.)
        }
    \label{fig:2D_H_WF}
\end{figure}

It is remarkable that only $6$ GWPs are required for decent accuracy throughout the propagation with the
greatest Rothe threshold $10^{-3}\,\text{a.u.}$, see Fig.~\ref{fig:2D_H_cost}. The highest accuracy, however, requires nearly an order of magnitude more GWPs in the basis at
the end of the simulation where the wavefunction complexity is substantial, as seen in Fig.~\ref{fig:2D_H_WF}.

\subsubsection{3D hydrogen atom}\label{sec:hydrogen_Rothe}

In Ref.~\cite{schraderTimeEvolutionOptimization2024}, Rothe's method is used to propagate a linear combination of isotropic GWPs for the hydrogen atom in a strong laser field. Due to the polarization of the laser along the $z$-direction and the resulting cylindrical symmetry of the time-dependent wave function (i.e., $m_l=0$), GWPs of the form
\begin{equation}
g_m(\mathbf{r}, t)
= g\bigl(\mathbf{r}; \boldsymbol{\alpha}_m(t)\bigr)
= N_m\exp\!\left(
- (a_m^2 + \ii b_m)\, \lVert \mathbf{r} - \boldsymbol{\mu}_m \rVert^2
+ \ii\, \mathbf{p}_m^{T} (\mathbf{r} - \boldsymbol{\mu}_m)
\right),
\end{equation}
are used.
Here, $\boldsymbol{\mu}_m = (0,0,\mu_m)^T$, $\mathbf{p}_m = (0,0,p_m)^T$ and $N_m$ is determined from the other parameters.

Due to the high variance in Coulomb potentials (see Sec. \ref{sec:Coulomb_variance}), the Hamiltonian is regularized,
\begin{equation} \label{eq:Hydrogen_Hamiltonian}
\hat{H}(t)
= -\frac{1}{2}\nabla^2
- \frac{\operatorname{erf}(100 r)}{r}
+ \hat{U}(t).
\end{equation}
Here, the matter-field coupling reads
\begin{equation}\label{eq:dipole_potential}
    \hat{U}(t)=E_0\, f(t)\, \sin(\omega t)\, z
\end{equation}
where $\omega=0.057\,\text{a.u.}$, $E_0\in\{0.03,0.06,0.12\}\,\text{a.u.}$, and 
\begin{equation}\label{eq:envelope}
f(t) = \sin^2\!\left(\frac{\pi t}{T_f}\right), \qquad 0 \le t \le T_f.
\end{equation}
Here, $T_f ={2\pi N_c}/{\omega}$ and $N_c=3$ is the number of optical cycles. We focus on $E_0=0.06\,\text{a.u.}$ here.

The ground state is represented using $25$ GWPs obtained by energy minimization. These are kept frozen, and the calculation starts with $25$ additional GWPs (i.e., in addition to the $25$ frozen ones) centered around the initial state. A GWP is added whenever the Rothe error crossed above $\varepsilon_{\Delta t}=\varepsilon/N_T$, where $N_T$ is the total number of time steps, and removed when it does not make the Rothe error greater them $\varepsilon_{\Delta t}$. A time step of $\Delta t=0.2\,\text{a.u.}$ is used.
Reference simulations are performed using the discrete variable representation (DVR).

\begin{figure}[htbp]
    \centering
    \includegraphics[width=\linewidth]{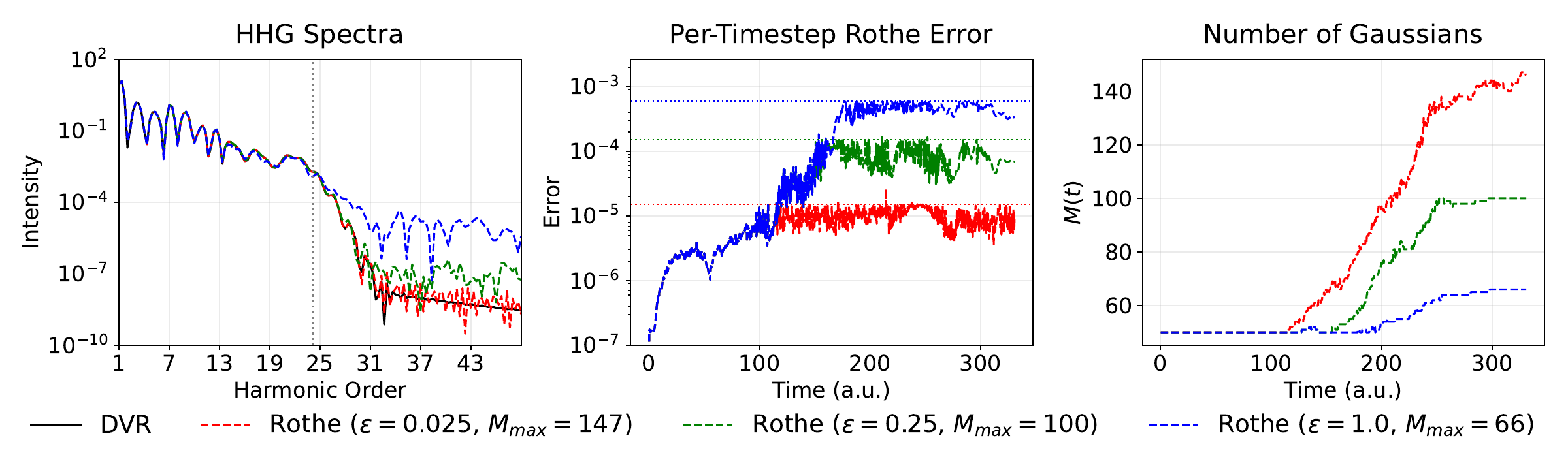}
    \caption{HHG spectra (left), per-timestep Rothe error (center), and number of GWPs $M(t)$ (right) for 3D hydrogen in a laser field with $E_0 = 0.06\,\text{a.u.}$, $\omega = 0.057\,\text{a.u.}$, and $N_c = 3$ optical cycles, computed using Rothe's method with three error thresholds $\varepsilon$ and compared to a converged DVR reference (black). The dotted vertical line in the HHG spectrum marks the classical cutoff energy at the $\sim 24$th harmonic, while the dotted lines in the per-timestep Rothe errors indicate the threshold $\varepsilon_{\Delta t}$.
    (Adapted from Ref.~\cite{schraderTimeEvolutionOptimization2024} with the permission of AIP Publishing.)
    }
    \label{fig:Hydrogen}
\end{figure}

Figure \ref{fig:Hydrogen} shows the extracted HHG spectra for $E_0=0.06\,\text{a.u.}$ using a Hann window function for different Rothe thresholds $\varepsilon\in\{0.025,0.25,1.0\}$, the Rothe error at each time step, and the number of GWPs at each time step. We observe that an increasing number of GWPs allows for lower Rothe errors, which in turn leads to better converged HHG spectra. Already with at most $66$ GWPs, the spectrum appears qualitatively correct. With at most $147$ GWPs, the HHG spectrum has essentially converged. This shows that one can represent a process as complicated as HHG in a strong field using less than $1000$ real parameters to represent the wave function, a compactness that reflects the adaptive basis tracking the wavefunction rather than covering a fixed spatial domain.

\subsubsection{Molecular systems}\label{sec:TDHF_Rothe}

Reference \cite{schraderTimeDependentGaussianBasis2025} applies Rothe's method to molecular electronic dynamics at the mean-field level (cf. Sec. \ref{sec:Rothe_MeanField}), using 1D models of $\mathrm{LiH}$ and $(\mathrm{LiH})_2$ \cite{SatoIshikawa2013} in a strong laser field. Both TDHF and TDDFT with an LDA functional \cite{Helbig2011,Wagner2012} are tested, with restricted Hartree-Fock and Kohn-Sham orbitals, respectively. Equation \eqref{eq:dipole_potential} is used for the matter-field interaction, with $\omega=0.06075\,\text{a.u.}$, $E_0\in\{0.0534,0.1068\}\,\text{a.u.}$, and $N_c=3$ optical cycles. To verify convergence, the density at the final time $\rho(x,t=T_f)$ as well as HHG spectra are compared to results from a converged grid calculation.

Basis functions are GWPs of the form
\begin{equation}
g_m({x}, t)
= g\bigl(x; \boldsymbol{\alpha}_m(t)\bigr)
= N_m \exp\!\left(
- (a_m^2 + \ii b_m)\, \left( x - {\mu}_m \right)^2
+ \ii {p}_m (x - {\mu}_m)
\right)
\end{equation}
where $N_m$ is a normalization constant. The initial orbitals at $t=t_0$ are determined by imaginary-time propagation; $24$ GWPs are sufficient for $\mathrm{LiH}$ (two spatial orbitals) and $38$ for $(\mathrm{LiH})_2$ (four spatial orbitals). A time step of $\Delta t=0.05\,\text{a.u.}$ is used in all cases.

\begin{figure}[htbp]
    \centering
    \includegraphics[width=\linewidth]{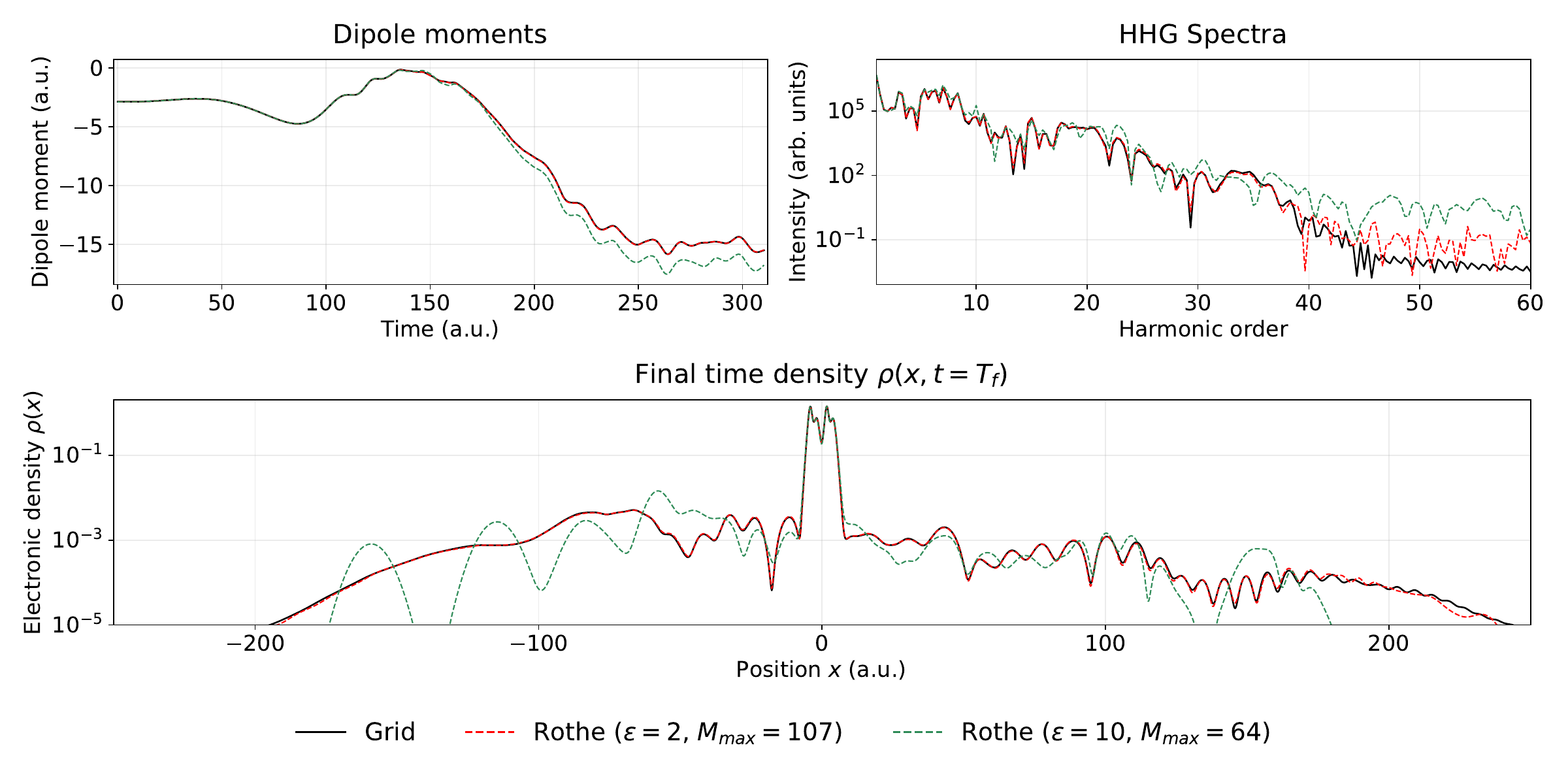}
    \caption{Dipole moment (top left), HHG spectrum (top right), and
final-time electronic density $\rho(x,t=T_f)$ (bottom) for
$(\mathrm{LiH})_2$ simulated with TDDFT/LDA at field strength
$E_0=0.0534\,\text{a.u.}$, frequency $\omega=0.06075\,\text{a.u.}$, and $N_c=3$ optical
cycles, comparing two Rothe runs with thresholds
$\varepsilon\in\{2,10\}\,\text{a.u.}$ (dashed, $M_\mathrm{max}\in\{107,64\}$
GWPs) against a converged grid reference (black).
(Adapted from Ref. \cite{schraderTimeDependentGaussianBasis2025}. Copyright  \copyright 2025 S. E. Schrader, H. E. Kristiansen, T. B. Pedersen, and S. Kvaal. Published by American Chemical Society.)
}
    \label{fig:TDDFT_LiH2_1}
\end{figure}
\begin{figure}[htbp]
    \centering
    \includegraphics[width=\linewidth]{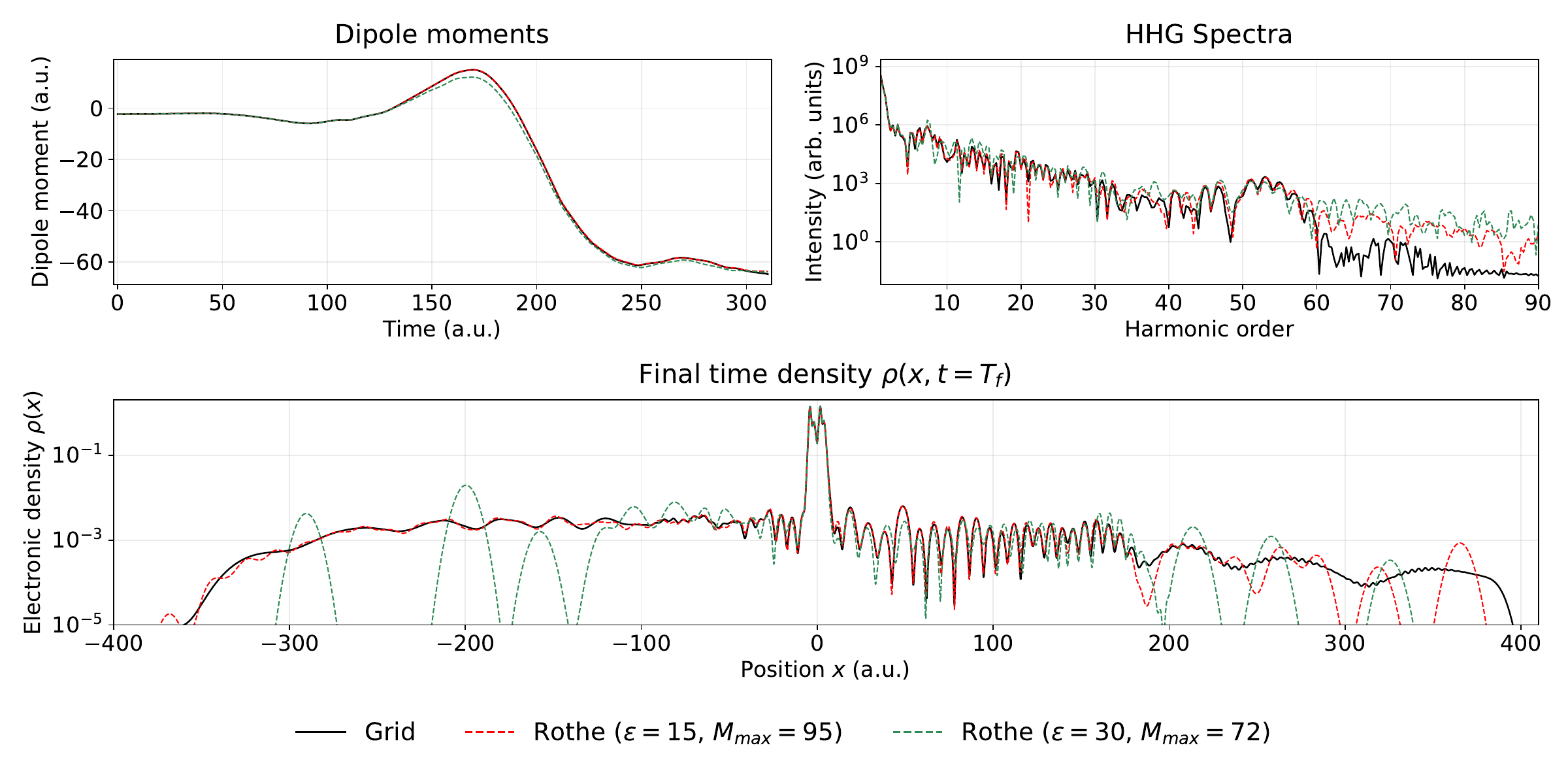}
    \caption{Dipole moment (top left), HHG spectrum (top right), and
final-time electronic density $\rho(x,t=T_f)$ (bottom) for
$(\mathrm{LiH})_2$ simulated with TDHF at the stronger field strength
$E_0=0.1068\,\text{a.u.}$, frequency $\omega=0.06075\,\text{a.u.}$, and $N_c=3$ optical
cycles, comparing two Rothe runs with thresholds
$\varepsilon\in\{15,30\}\,\text{a.u.}$ (dashed, $M_\mathrm{max}\in\{95,72\}$
Gaussians) against a converged grid reference (black).
(Adapted from Ref. \cite{schraderTimeDependentGaussianBasis2025}. Copyright  \copyright 2025 S. E. Schrader, H. E. Kristiansen, T. B. Pedersen, and S. Kvaal. Published by American Chemical Society.)
}
    \label{fig:TDHF_LiH2_4}
\end{figure}

Figures \ref{fig:TDDFT_LiH2_1} and \ref{fig:TDHF_LiH2_4} show the dipole moment, HHG spectrum, and density at $t=T_f$ for the $(\mathrm{LiH})_2$ dimer, simulated using TDDFT at a field strength of $E_0=0.0534\,\text{a.u.}$, and using TDHF at a field strength of $E_0=0.1068\,\text{a.u.}$, respectively. We observe an improvement of the quality of the dipole moments, the HHG spectra, and the density with an increasing number of GWPs at both field strengths considered. For the weaker field strength (Fig.~\ref{fig:TDDFT_LiH2_1}), with the larger number of GWPs considered, the data is essentially converged. This is despite a very large upper bound of the Rothe error of $\varepsilon=2\,\text{a.u.}$, which shows that the magnitude of the Rothe error does not necessarily imply the quality of both time-averaged quantities (such as a HHG spectrum) and the quality of the resulting wave function at $t=T_f$. For the stronger field (Fig.~\ref{fig:TDHF_LiH2_4}), we observe that the dipole moment, HHG spectrum and density are not fully converged with $95$ GWPs. However, the HHG spectrum is still quantitatively correct up to the $\sim 60$th harmonic, and the density is quantitatively correct in the range $[-300,200]\,\text{a.u.}$. Although $\varepsilon=15\,\text{a.u.}$ is large enough that one would naively expect the dynamics to be unreliable, the actual agreement with the reference is good, and the marked improvement from $\varepsilon=30\,\text{a.u.}$ to $\varepsilon=15\,\text{a.u.}$ confirms that $\varepsilon$ is a meaningful target for optimization.

\subsection{Rovibrational dynamics}

\subsubsection{2D Morse potential}

Turning to nuclear motion, it is of interest to investigate the performance of Rothe's method with GWPs that are free to move in Hilbert space without the local harmonic approximation.
Refence \cite{wozniak2025rothetimepropagationcoupled} presents a simple test of rovibrational dynamics in the 2D Morse potential,
\begin{equation}
    V(x,y) = D_e \left[ \mathrm{e}^{-2\alpha (\sqrt{x^2+y^2} - r_e)} - 2\mathrm{e}^{-\alpha (\sqrt{x^2+y^2} - r_e)} \right],
\end{equation}
with $D_e = 0.17449\,\text{a.u.}$, $r_e = 1.4011\,\text{a.u.}$, and $\alpha = 1.4556\,\text{a.u.}$. For simplicity, a single particle with mass
$1605.487\,\text{a.u.}$ and unit charge is assumed to be initially in the ground state of the Morse potential. The ground-state wavefunction is computed variationally
as a linear combination of $8$ GWPs centered at the origin with zero momentum. The GWPs are allowed to have complex diagonal and isotropic covariance matrix. The imaginary
part of the covariance is needed to correctly describe the annular shape of the ground-state wavefunction, peaking at distance $r_e$ from the origin. The dynamics are driven
by a very strong electric-field impulse along the $x$-axis. The explicit form of the electric field is that of Eq.~\eqref{eq:electric field 1D} with $\omega=0\,\text{a.u.}$,
$E_0 = 2\,\text{a.u.}$, $t_0 = 0\,\text{a.u.}$, and $t_1 = 20\,\text{a.u.}$. This impulse briefly but violently pushes the particle in the positive $x$-direction.

\begin{figure}[htbp]
    \centering
    \includegraphics[width=\textwidth]{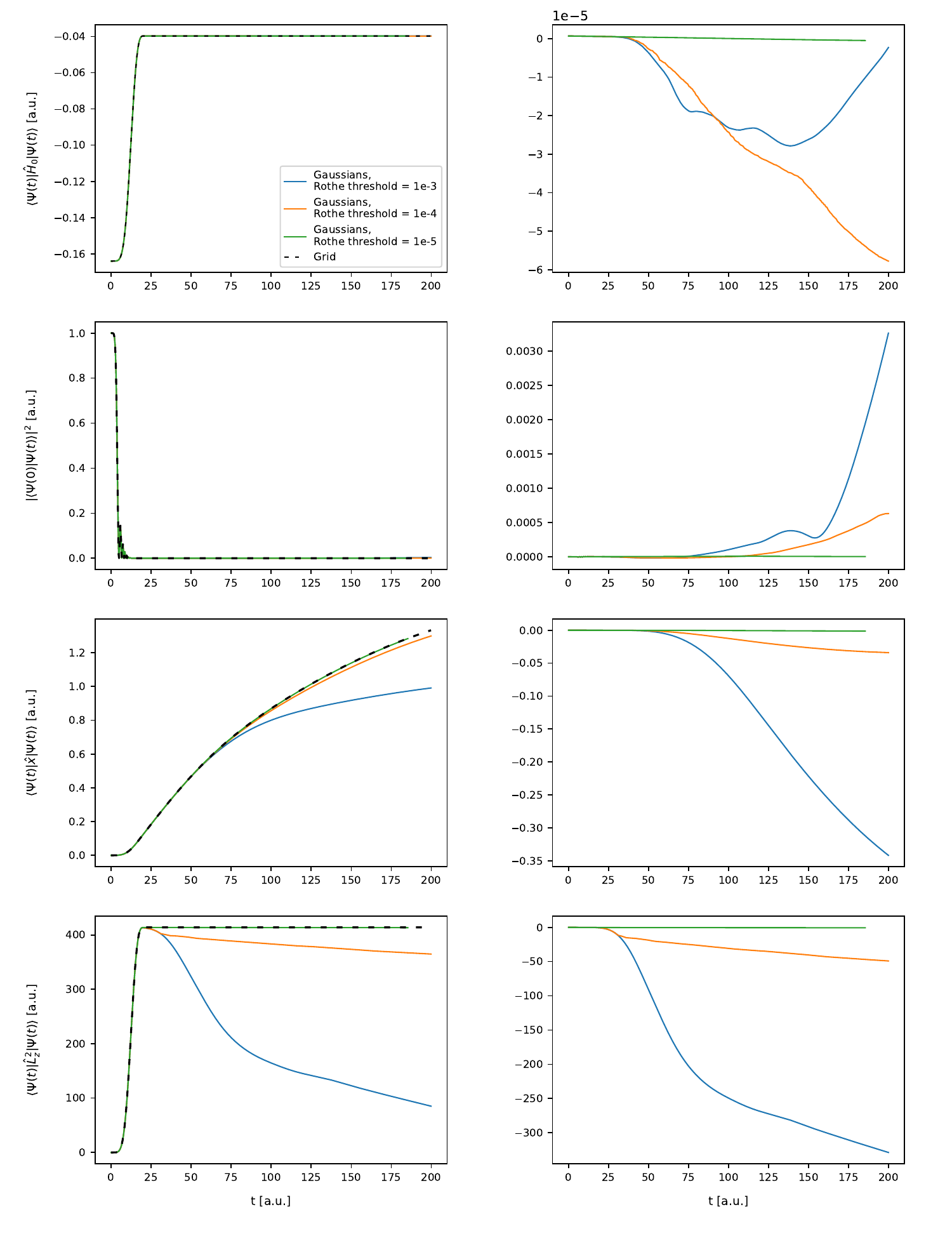}
    \caption{Left column: expectation values as functions of time for three values the Rothe minimization threshold along with the grid-based reference results for
        the 2D Morse potential.
        Right column: Errors with respect to the reference expectation values. Rows from top to bottom contain results for the energy, ground-state survival probability,
        average position on the $x$-axis, and squared angular momentum, respectively.
        (Reprinted from Ref. \cite{wozniak2025rothetimepropagationcoupled}. Copyright  \copyright 2025 A. P. Wo{\'z}niak, L. Adamowicz, T. B. Pedersen, and S. Kvaal. Published by American Chemical Society.)
        }
    \label{fig:2D_Morse_observables}
\end{figure}

The wavefunction is propagated from $t=0\,\text{a.u.}$ to $t=200\,\text{a.u.}$ using Rothe's method with time step $\Delta t = 0.05\,\text{a.u.}$ and the same three Rothe
thresholds used for the 2D hydrogen atom above ($10^{-3}$, $10^{-4}$, $10^{-5}\,\text{a.u.}$).
Also the same observables are recorded and compared with a highly accurate grid reference simulation, as shown in
Fig.~\ref{fig:2D_Morse_observables}.
The electric-field impulse excites the particle to a superposition of states with an energy close to the dissociation limit.
After the impulse, energy and ground-state survival probability are reasonably well conserved at any of the three thresholds,
with noticeable improvements as the Rothe threshold is lowered.
The position ($x$) expectation value, however, requires tight thresholds to reach errors below a few percent. The same can be observed for the
squared angular momentum, which should be conserved after the impulse. Comparing with the 2D hydrogen simulations above, it would seem that the rovibrational
dynamics induced by the electric-field impulse in the Morse potential is somewhat more difficult to capture correctly with only a few GWPs. 

\begin{figure}[htbp]
    \centering
    \includegraphics[width=\textwidth]{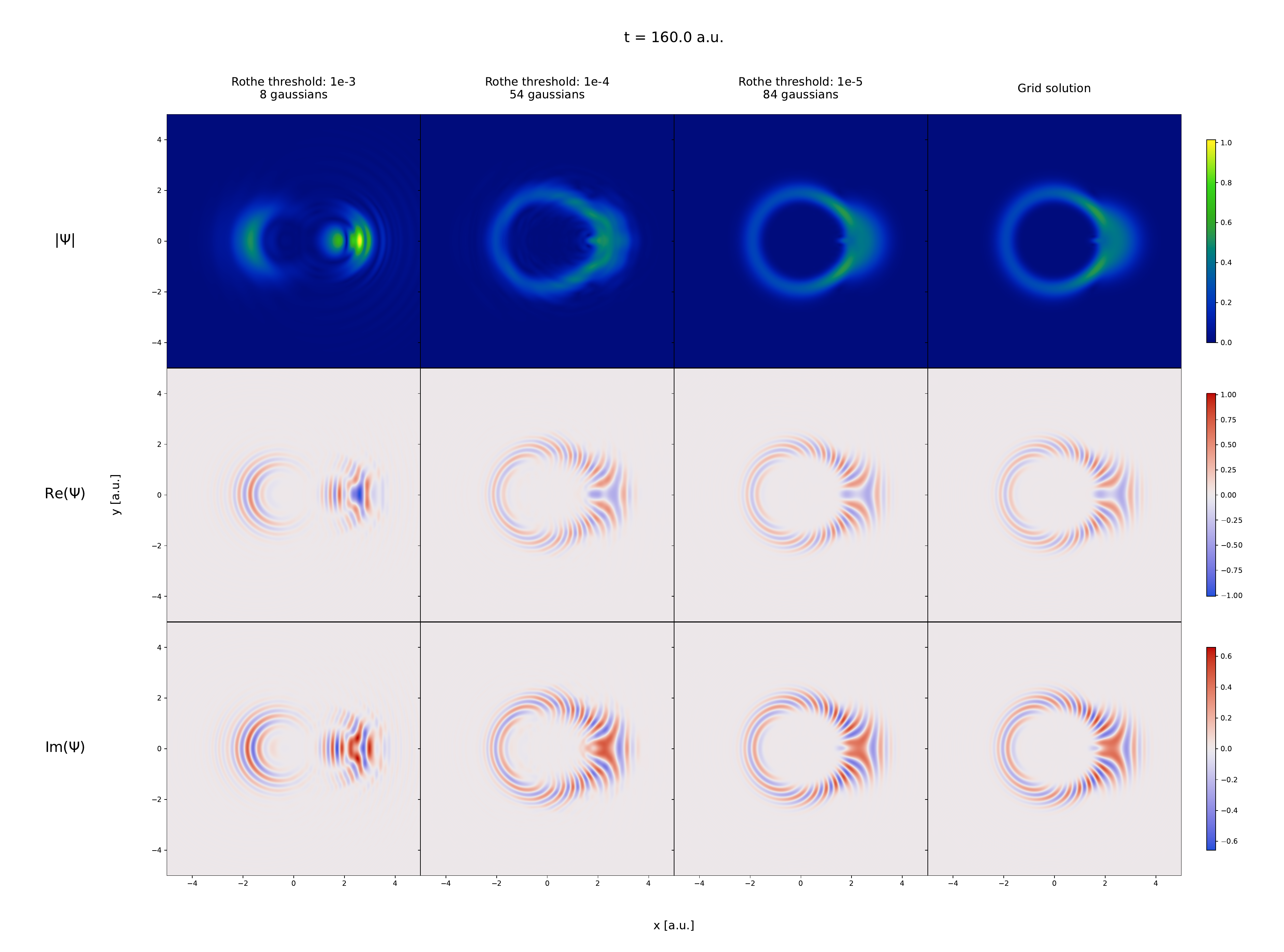}
    \caption{Plot of the absolute value and the real and imaginary parts of the 2D Morse wavefunction at $t=160\,\text{a.u.}$ for three different Rothe thresholds.
        The last column shows the reference wavefunction.
        (Reprinted from Ref. \cite{wozniak2025rothetimepropagationcoupled}. Copyright  \copyright 2025 A. P. Wo{\'z}niak, L. Adamowicz, T. B. Pedersen, and S. Kvaal. Published by American Chemical Society.)
        }
    \label{fig:2D_Morse_WF}
\end{figure}

Indeed, as seen in the snapshots at $t=160\,\text{a.u.}$ in Fig.~\ref{fig:2D_Morse_WF}, the wavefunction is quite far from converged with the threshold $10^{-3}\,\text{a.u.}$,
which only requires $8$ GWPs.
Reducing the threshold to $10^{-4}\,\text{a.u.}$, an additional $46$ GWPs must be added ($54$ in total), leading to substantial improvement of the
wavefunction. Compared with the grid reference, however, artifacts are still clearly visible, especially at positive $x$ where spurious interference patterns appear.
These are essentially removed by further reducing the threshold to $10^{-5}\,\text{a.u.}$, at the cost of $30$ additional GWPs ($84$ in total).

\subsubsection{Henon-Heiles potential}

The $D$-dimensional Henon-Heiles potential is an anharmonic potential giving rise to complicated dynamics, including classical chaotic behavior. It is often used to test
the accuracy of quantum dynamics methods \cite{Burghardt_Worth_HeHe}, and is employed for testing the Rothe method in Ref.~\cite{schraderMultidimensionalQuantumDynamics2025} with
a single Gaussian as initial state.
The $D$-dimensional Hamiltonian reads in natural units
\begin{align}
    \hat{H} &=-\frac{1}{2} \sum_{i=1}^D \frac{\partial^2}{\partial r_i^2}+\frac{1}{2} \sum_{i=1}^D r_i^2 \nonumber
    +\lambda \sum_{i=1}^{D-1}\left(r_i^2 r_{i+1}-\frac{1}{3} r_{i+1}^3\right), \label{eq:hamiltonian}
\end{align}
where the last term couples different degrees of freedom, making the dynamics anharmonic and complicated. Following Ref. \cite{Burghardt_Worth_HeHe}, $\lambda=0.111803$.

To properly describe the coupled degrees of freedom, the time-dependent wavefunction is represented as a linear combination of complex ECGs (Eq.~\eqref{eq:multidimensional gaussian}),
with the initial state is given by a single uncorrelated Gaussian displaced from the origin by $2$ in each direction,
\begin{equation}\label{eq:InitialGaussian}
\Psi(\boldsymbol{r},t=0)=\pi^{-D/4} \exp\left(-\left(\boldsymbol{r}-\boldsymbol{\mu}^0\right)^T\boldsymbol{A}^0\left(\boldsymbol{r}-\boldsymbol{\mu}^0\right)\right).
\end{equation}
Here, $\boldsymbol{A}^0=\frac{1}{2}\boldsymbol{I}_D$ and $\mu^0_i=2$ for $i=1,\dots,D$. The wavefunction is propagated in $10\,000$ time steps, each with $\Delta t=0.01\,\text{a.u.}$ (i.e., $T_f = 100\,\text{a.u.}$).
The target quantity is the dampened spectrum defined as
\begin{equation}
    \label{eq:spectrum}
    S(\omega)\propto \operatorname{Re} \int_0^{\infty} \exp\left(\ii \omega t-t/30\right) C(t) d t,
\end{equation}
where $C(t)=\langle \Psi(0)|\Psi(t)\rangle$ is the autocorrelation function, which for real initial states can be conveniently calculated as 
\begin{equation}
    C(t)= \langle \Psi^*(t/2)|\Psi(t/2)\rangle.
\end{equation}
\begin{figure}[htbp]
    \centering
    \includegraphics[width=\linewidth]{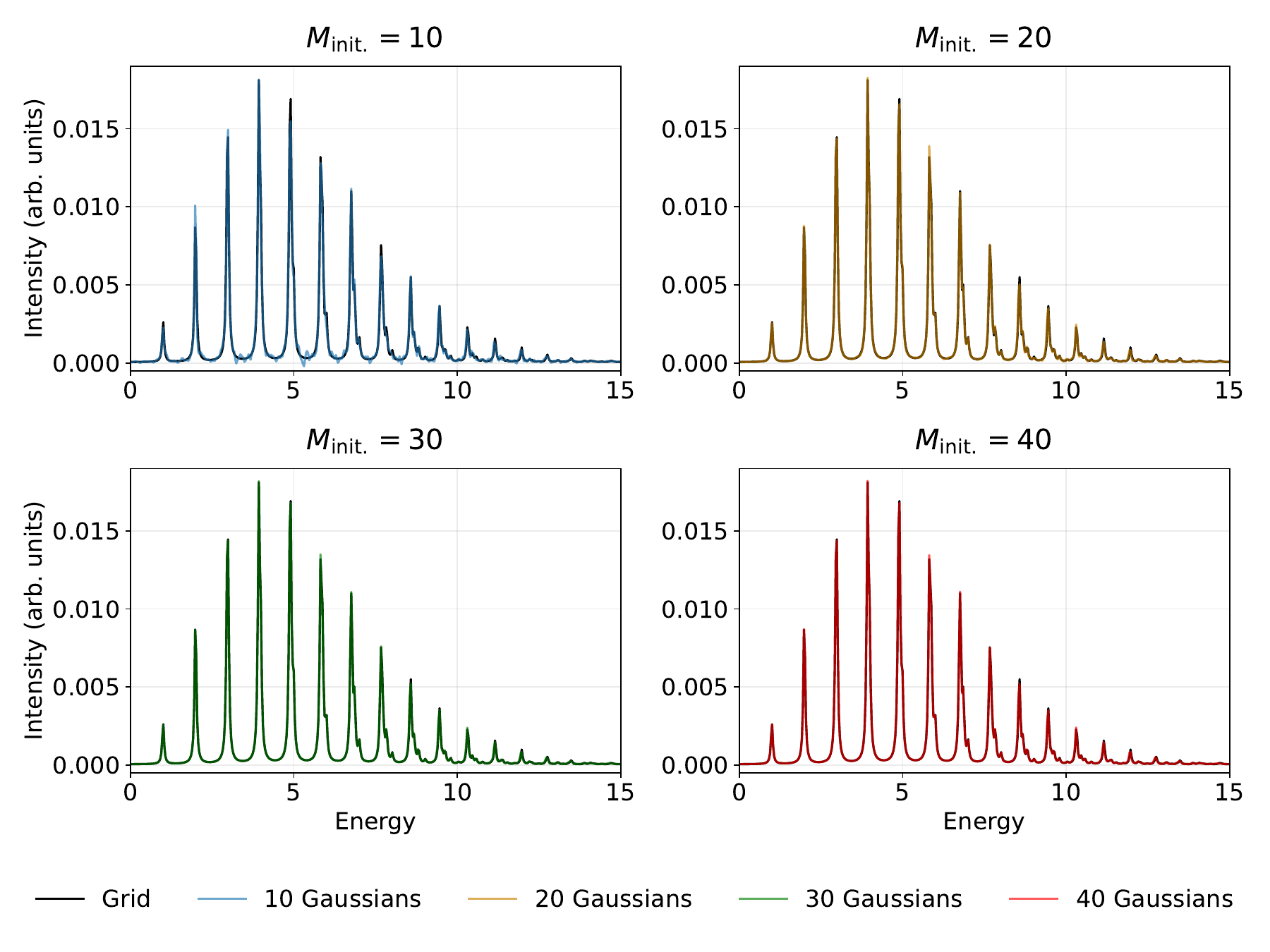}
    \caption{Spectra $S(\omega)$ (Eq.~\eqref{eq:spectrum}) for the 2D Henon-Heiles potential with $M_\text{init}\in\{10,20,30,40\}$ initial Gaussians (colored lines), compared to essentially exact grid calculation (black line).
(Adapted from Ref.~\cite{schraderMultidimensionalQuantumDynamics2025} with the permission of AIP Publishing.)
}
    \label{fig:HeHe_2D}
\end{figure}
\begin{figure}[htbp]
    \centering
    \includegraphics[width=\linewidth]{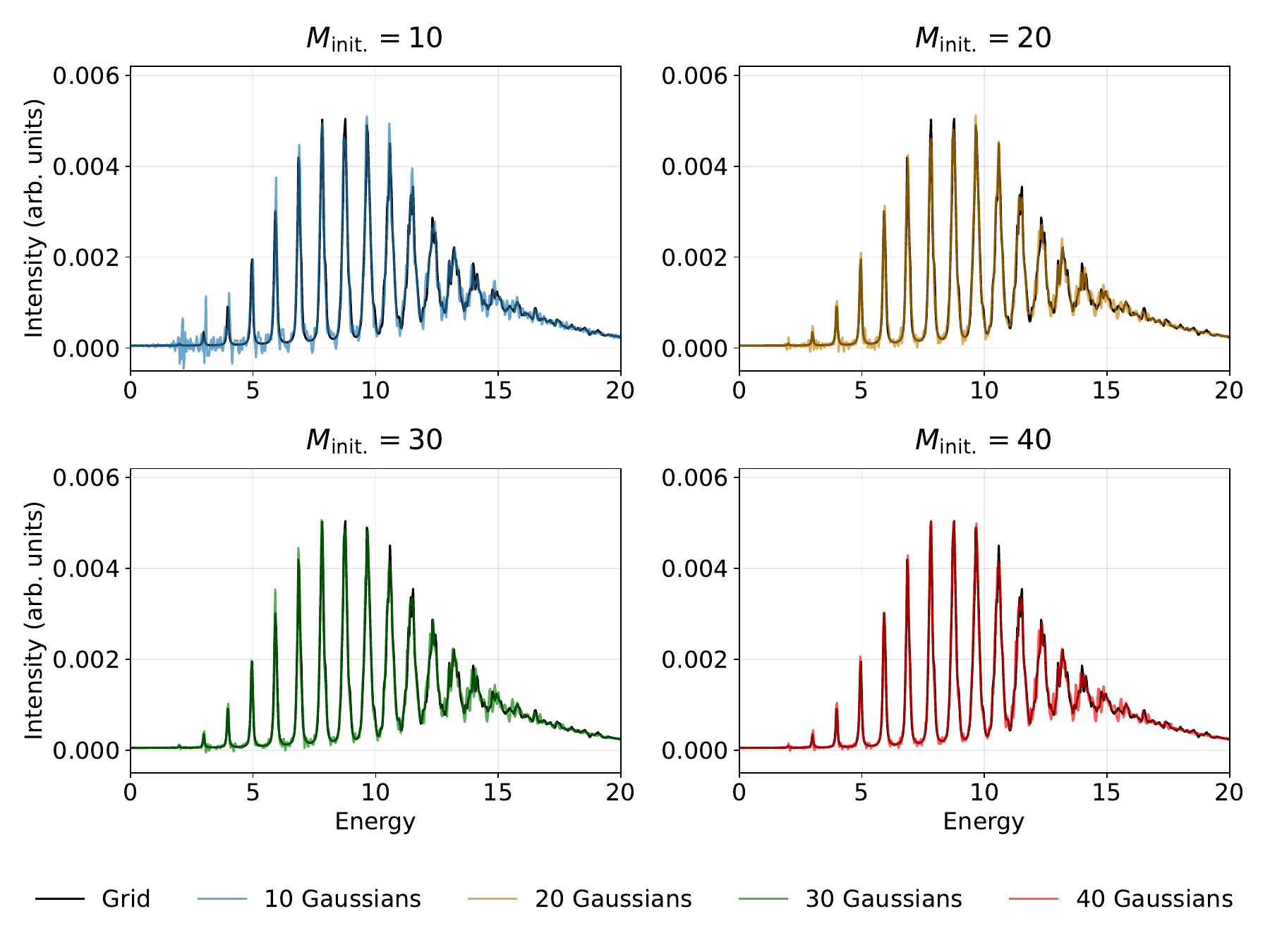}
    \caption{Spectra $S(\omega)$ (Eq.~\eqref{eq:spectrum}) for the 4D Henon-Heiles potential with $M_\text{init}\in\{10,20,30,40\}$ initial Gaussians (colored lines), compared to essentially exact grid calculation (black line).
(Adapted from Ref.~\cite{schraderMultidimensionalQuantumDynamics2025} with the permission of AIP Publishing.)
    }
    \label{fig:HeHe_4D}
\end{figure}
The number of ECGs is kept fixed throughout the Rothe propagation. I.e., no ECGs were added, although some outgoing ECGs were removed due to the unboundedness of the potential.
Thus, the simulation starts with $M-1$ ECGs added to the initial single Gaussian. This is done because the Rothe error becomes so large in the Henon-Heiles potential that the
number of ECGs would explode with a sufficiently low Rothe threshold. The additional ECGs are clustered around the initial state, Eq.~\eqref{eq:InitialGaussian}, and become populated
in the course of the propagation.

Spectra are computed for $D\in\{2,3,4\}$, $M_{init}\in\{5,10,20,30,40\}$ and compared with an essentially exact grid calculation.
Results for $D=2$ and $D=4$ are shown in Figs. \ref{fig:HeHe_2D} and \ref{fig:HeHe_4D}, respectively, for $M_{init}\in\{10,20,30,40\}$.
In the 2D case, we observe that already with $20$ ECGs, the spectrum is essentially converged and only minor improvements are obtained by increasing the basis size.
In the 4D case, on the other hand, we find that the spectrum is essentially, but not fully, converged with $40$ ECGs.

\section{Concluding remarks}
\label{section: conclusion}

Time-dependent GWPs form a highly adaptive yet compact basis for simulations of atomic and molecular quantum dynamics but are, unfortunately,
exceedingly challenging to propagate in a numerically stable manner for potentials that are not harmonic.
In this work, we have described the underlying
sources of these challenges and reviewed Rothe's method as an alternative to conventional time-dependent variational principles.
Using simplified model systems representing different aspects of the molecular Coulomb Hamiltonian, numerical tests indicate that
Rothe's method does indeed provide a viable path toward black-box adaptive propagation of linear combinations of GWPs, yielding
simulation results on par with much more memory-consuming grid-based methods such as the discrete variable representation.
The results highlight that Rothe's method allows for the propagation of correlated and uncorrelated Gaussians in a variety of contexts,
including nuclear quantum dynamics and electron dynamics. Rothe's method can be applied to both the time-dependent Schrödinger equation directly
and to orbital equations arising in methods such as time-dependent density-functional theory.
Challenges within Rothe's method have been pointed out, including the need for matrix elements of the squared Hamiltonian
and the issue of conservation laws which are only strictly fulfilled when the Rothe optimization problem is solved exactly.
We have also pointed out that the difficulty of capturing wavefunction cusps with a Gaussian basis is emphasized by the Rothe method
which, to leading order for time-independent Hamiltonians, depends explicitly on minimization of the energy variance.

Without disregarding the importance of computational efficiency for practical applications of Rothe's method, our numerical tests have been
conducted using preliminary implementations (e.g., quadrature-based calculations of matrix elements) and, therefore, computational timings have not been discussed.
To make Rothe's method widely applicable for quantum-dynamics studies of realistic systems, developing an efficient software library for the calculation of the squared Hamiltonian and its derivatives for complex-valued, floating Gaussians is a necessary prerequisite.

To further extend Rothe's method, we see several directions. The first one is algorithmic improvements. In particular, in Ref. \cite{schrader2026rothe}, we suggest a regularization scheme that \emph{forces} an approximate ground state $\ket{\Psi_a}$ to be the exact ground state of a regularized Hamiltonian $\hat{H}_a$, defined as
\begin{equation}
    \hat{H}_a=\hat{H}+\hat{V}_{a},
\end{equation}
where
\begin{equation}
    \hat{V}_a=(E_a-\hat H)|\Psi_a\rangle\langle \Psi_a| + |\Psi_a\rangle\langle \Psi_a|(E_a-\hat{H}),
\end{equation}
with $E_a=\bra{\Psi_a}\hat H\ket{\Psi_a}$. This scheme ensures that the energy variance of the initial state is exactly zero, and one no longer needs to use excessive numbers of Gaussians to very accurately represent, e.g., cusps of the ground state. Regularization of singular potentials will also not be needed. Initial results presented in Ref. \cite{schrader2026rothe} indicate that the resulting bias in the dynamics is small.

A second path along which one can exploit the power of Rothe's method is within computational models such as MCTDH(F) \cite{beck2000multiconfiguration, Meyer2009, hochstuhl_time-dependent_2014, lode2020colloquium} and orbital-adaptive or orbital-optimized time-dependent coupled-cluster theory \cite{kvaal2012ab,sato2018communication,ofstad_time-dependent_2023,hojlund2024time, hojlund2024bivariational} where the orbital equations of motion can be recast as optimization problems. This would allow the use of Gaussian basis sets to study the dynamics of molecules using correlated, orbital-based methods. 

Finally, Rothe's method can pave the way for highly accurate non-Born-Oppenheimer simulations using fully flexible many-body ECGs for small molecular systems,
providing benchmarks for developing improved computational models for larger molecular systems and new phenomena. Such simulations would also offer fundamental insights into the
limits of validity of the Coulomb Hamiltonian and semiclassical matter-field interactions for atomic and molecular systems.

\section*{Acknowledgements}

This work was supported by the Research Council of Norway through its Centres of Excellence scheme, project no.~262695.
Support from the National Science Foundation, grant no.~1856702, is also acknowledged.
The calculations presented in this work were carried out using resources provided
by Sigma2---the National Infrastructure for High Performance Computing and Data Storage in Norway, grant no.~NN4654K,
by University of Arizona Research Computing, and
by Wroclaw Centre for Networking and Supercomputing, grant no.~567.

\printbibliography{}

\end{document}